\documentclass[twocolumn]{aastex63}
\usepackage{graphics,epsf}
\usepackage{amsmath}                
\usepackage{amsfonts}               
\usepackage{amssymb}                
\usepackage{epsfig}                 
\usepackage{appendix}
\usepackage{graphicx}
\usepackage{float}
\usepackage{color}
\usepackage{multirow}
\usepackage{colortbl}
\usepackage[para,online,flushleft]{threeparttable}

\hypersetup{citecolor=blue, 
            linkcolor=red, 
            menucolor=blue, 
            urlcolor=blue}  

\newcommand{\cm}{{~\rm cm}}
\newcommand{\m}{{~\rm m}}
\newcommand{\km}{{~\rm km}}
\newcommand{\s}{{~\rm s}}
\newcommand{\h}{{~\rm h}}

\newcommand{\g}{{~\rm g}}
\newcommand{\G}{{~\rm G}}
\newcommand{\K}{{~\rm K}}
\newcommand{\erg}{{~\rm erg}}
\newcommand{\yr}{{~\rm yr}}

\newcommand{\AU}{{~\rm AU}}

\begin{document}

\title{Simulating the deposition of angular momentum by jets in common envelope evolution}

\author{Ron Schreier}
\affiliation{Department of Physics, Technion, Haifa, 3200003, Israel; \\ 	
ronsr@physics.technion.ac.il, shlomi.hillel@gmail.com, soker@physics.technion.ac.il}

\author{Shlomi Hillel}
\affiliation{Department of Physics, Technion, Haifa, 3200003, Israel; \\ 	
ronsr@physics.technion.ac.il, shlomi.hillel@gmail.com, soker@physics.technion.ac.il}

\author[0000-0003-0375-8987]{Noam Soker}
\affiliation{Department of Physics, Technion, Haifa, 3200003, Israel; \\ 	
ronsr@physics.technion.ac.il, shlomi.hillel@gmail.com, soker@physics.technion.ac.il}

\begin{abstract}

We conducted three-dimensional hydrodynamical simulations of common envelope evolution (CEE) of a neutron star (NS) or a black hole (BH) inside a red supergiant (RSG) envelope and find that the jets that we expect the NS/BH to launch during the CEE spin-up the common envelope. We find that when the NS/BH launches jets that are exactly perpendicular to the orbital plane (the jets are aligned with the orbital angular momentum) the jets deposit angular momentum to the envelope that is aligned with the orbital angular momentum. When the jets' axis is inclined to the orbital angular momentum axis so is the angular momentum that the jets deposit to the envelope. Such tilted jets might be launched when the NS/BH has a close companion when it enters the RSG envelope. We did not allow for spiralling-in and could follow the evolution for only three orbits. The first orbit mimics the plunge-in phase of the CEE, when the NS/BH rapidly dives in, while the third orbit mimics the self-regulated phase when spiralling-in is very slow. We find that the jets deposit significant amount of angular momentum only during the plunge-in phase. A post-CEE core collapse supernova explosion will leave two NS/BH, bound or unbound, whose spin might be misaligned to the orbital angular momentum. Our results strengthen an earlier claim that inclined-triple-star CEE might lead to spin-orbit misalignment of NS/BH-NS/BH binary systems. 
\end{abstract}

\keywords{(stars:) binaries (including multiple): close; (stars:) supernovae: general; transients: supernovae; stars: jets} 

\section{Introduction} 
\label{sec:intro}

Common envelope jets supernovae (CEJSNe) are weeks to years optical transients of binary and multiple stellar systems that experience the common envelope evolution (CEE) of a neutron star (NS) or a black hole (BH) inside the envelope and then the core of a red supergiant (RSG) star. As the NS/BH spirals-in inside the envelope and then inside the core of the RSG star it accretes mass with sufficiently large specific angular momentum to form an accretion disk around the NS/BH. The accretion disk launches powerful jets to power the CEJSN (e.g., \citealt{Gilkisetal2019, SokeretalGG2019, GrichenerSoker2019a, Schroderetal2020, GrichenerSoker2021, Hilleletal2022FB}). The accurate definitions of a CEJSN event is a CEE where the NS/BH enters the core at the end of the CEE or destroys the core by tidal interaction. On the other hand, if the RSG core stays intact and the powering is only due to accretion from the RSG envelope the event is termed CEJSN-impostor. In this study we will use the term CEJSN for both cases. 

The compactness of the NS/BH enables three essential processes for a CEJSN event to take place. (1) The formation of an accretion disk despite the relatively small specific angular momentum of the gas that the NS/BH accretes from the inhomogeneous envelope or from the core of the RSG star (e.g.,  \citealt{ArmitageLivio2000, Papishetal2015, SokerGilkis2018,  LopezCamaraetal2019, LopezCamaraetal2020MN}; see \citealt{Hilleletal2022FB} and \citealt{LopezCamaraetal2022MS} for further discussion and references). 
(2) An efficient neutrino-cooling close to the NS/BH that reduces the energy of the accreted mass and by that enables a very high mass accretion rate $\dot M_{\rm acc}$ (\citealt{HouckChevalier1991, Chevalier1993, Chevalier2012}). Jets carry the rest of the energy, and a BH can accrete part of the energy (e.g., \citealt{Pophametal1999}). The accretion rate in CEJSNe is $\dot M_{\rm acc} \ga 10^{-3} M_\odot \yr^{-1}$, the minimum required for an efficient neutrino cooling. 
(3) The deep potential well implies a very large amount of accretion energy per unit mass, and the possibility of relativistic jets. Production of energetic neutrinos in the case of a BH launching jets in the envelope \citep{GrichenerSoker2021} and r-process nucleosynthesis when a NS accretes from the core of a RSG (e.g., \citealt{GrichenerSoker2019a}) might take place in such jets. 

In cases where the NS/BH ejects the RSG envelope before it reaches the core (e.g., \citealt{SokeretalGG2019}) it leaves the massive core to explode later as a core collapse supernova (CCSN), which might be the second CCSN, or event the third CCSN in some triple-star systems. The end product is a NS/BH - NS/BH system, bound or unbound. In cases where the  NS/BH continues to spiral-in to the core it destroys the core. The core material forms a massive accretion disk that in turn launches very energetic jets (e.g., \citealt{GrichenerSoker2019a}) with energies of up to $\simeq {\rm several} \times 10^{52} \erg$. The transient event can be classified as a super luminous supernova (SLSN) with a circumstellar matter (CSM), but it is a CEJSN event rather than a CCSN. The long-lasting light curve might have several peaks (e.g., \citealt{SokeretalGG2019, Schroderetal2020}). \cite{SokerGilkis2018} suggest that such events might account for the enigmatic SN~iPTF14hls (observations by \citealt{Arcavietal2017}) and similar transients, e.g., SN~2020faa (observations by \citealt{Yangetal2021}). A different type of CEJSN-impostor event \citep{Soker2022FBOT} or a different type of CEJSN \citep{SokeretalGG2019} might account for fast-rising blue optical transients, e.g., AT2018cow (e.g., \citealt{Prenticeetal2018, Marguttietal2019, NayanaChandra2021}), AT2020xnd (observations by \citealt{Perleyetal2021}), and AT2020mrf (observations by \citealt{Yaoetal2022}).
   
Because the jets unbind a large fraction of the envelope mass they increase the CEE efficiency parameter, possibly to values $\alpha_{\rm CE}>1$, as some scenarios requires (e.g. \citealt{Fragosetal2019, BroekgaardenBerger2021,  Garciaetal2021, Zevinetal2021}). 
In addition, the merger of the NS/BH with the core is a gravitational waves source (e.g., \citealt{Ginatetal2020}). 

\begin{figure}[t!]
\includegraphics[trim=5.5cm 5.3cm 0.0cm 2.0cm ,clip, scale=0.55]{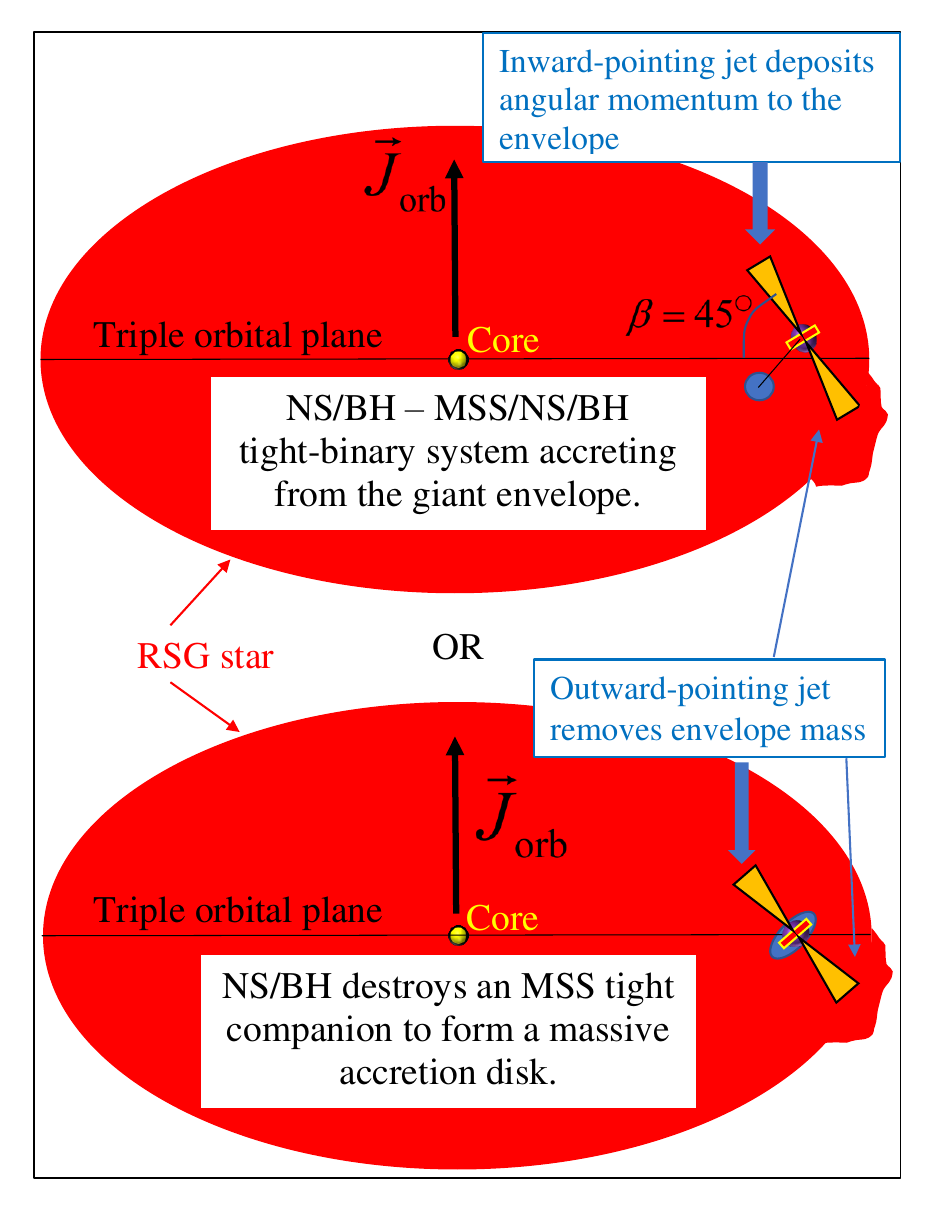}
\caption{A schematic diagram of the two types of triple-star interactions that might lead to tilted jets (adopted from \citealt{Soker2022misalignment}). A tight binary system spirals-in inside the envelope of an RSG. In the upper panel accretion from the RSG envelope forms the accretion disk around one of the tight binary stars or a disk around each of the two tight binary stars. The jets'-axis is determined by both the orbital angular momenta of the  tight binary system and that of the triple system. In the lower panel a NS/BH destroys its close main sequence companion in the tight binary system to form an accretion disk that is tilted to the triple-star orbital plane.  
Abbreviation: BH: black hole; MSS: main sequence star; NS: neutron star; RSG: red supergiant. }
 \label{fig:Schamnatic}
 \end{figure}

In a recent paper one of us \citep{Soker2022misalignment} proposes that jets in a triple-star CEE can deposit angular momentum to the giant envelope with components at large angles ($> 10^\circ$)  to the orbital angular momentum axis of the triple-star system.    
We schematically present the two types of triple-star interactions that might lead to tilted jets in Fig. \ref{fig:Schamnatic} that we adopt from \cite{Soker2022misalignment}.
According to the scenario, this might occur if the orbital plane of the tight binary system is inclined to the orbital plane of the triple system, such that one or two stars of the tight binary system launch jets that are inclined to the angular momentum axis of the triple system, i.e., the orbital angular momentum of the orbit of the tight binary system and the RSG around their mutual center of mass.

In cases where the tight binary system contains a NS/BH or two and the giant is a RSG that later collapses to form a NS/BH, the spin axis of this newly born NS/BH will be misaligned with the orbital angular momentum . This CEJSN scenario might lead to spin-orbit misalignment through a CEE. \cite{Soker2022misalignment} discusses more triple-star evolutionary channels that might lead to spin-orbit misalignment of NS/BH-NS/BH binary systems. 
Triple star CEJSN evolution (e.g., \citealt{Soker2021Double, Soker2021NSNSmerger, AkashiSoker2021}) might not be so rare because of the very large fraction of massive triple-star systems (e.g., \citealt{Sanaetal2014, MoeDiStefano2017}) and their interaction outcomes (e.g., \citealt{Toonenetal2021}). There are other studies of triple star CEE that do not necessarily include NS/BH (e.g., \citealt{SabachSoker2015, Hilleletal2017, Schreieretal2019inclined, ComerfordIzzard2020, GlanzPerets2021, SokerBear2021Parasite}).

We set to hydrodynamically simulate the deposition of misaligned angular momentum by jets in CEJSNe. In section \ref{sec:Numerical} we describe the numerical code and setting. In section \ref{sec:FlowStructure} we present the flow structure resulting from our simulations. In section \ref{sec:Perp} we describe the results of simulating jets perpendicular to the orbital plane, i.e., jets along the orbital angular momentum axis, and in section \ref{sec:Tilted} we describe the results for jets that are inclined to the orbital angular momentum axis.
 As far as we know this study is the first one of angular momentum deposition by jets into a common envelope. We therefore isolate the effects of the jets from the spiralling-in and the NS/BH gravity. Namely, we do not include spiralling-in nor the NS/BH gravity.   In section \ref{sec:Reliability} we discuss the reliability and implications of our work. We summarize our results in section \ref{sec:Summary}.

\section{The numerical setup}
\label{sec:Numerical}

We begin by using the results of a stellar evolution simulation using \texttt{MESA} \citep{Paxtonetal2011, Paxtonetal2013, Paxtonetal2015, Paxtonetal2018, Paxtonetal2019}.
The simulation consists of a zero-age-main-sequence star of mass $M_{\rm 1,ZAMS}=15 M_\odot$ and metallicity of $Z=0.02$, evolving for $1.1\times10^6 \yr$, becoming a RSG star with a radius of $R_{\rm RSG}=881\,R_{\odot}$, a mass of $M_1=  12.5 M_\odot$, and an effective temperature of $T_{\rm eff}= 3160K$. 
This stellar model is then transferred to a three-dimensional (3D) hydrodynamical code {\sc flash} \citep{Fryxelletal2000} and placed at the center of the computational grid.
The composition of the 3D star is pure hydrogen, and the code assumes full ionisation. Outside the stellar model, the numerical grid is filled with gas of very low density $\rho_{\rm grid,0} = 2.1 \times 10^{-13} \g \cm^{-3}$ and of temperature $T_{\rm grid,0}= 1100 \K$.  
For numerical reasons we also replace the inner $20\%$ of the stellar radius, $R_{\rm in} = 176\,R_{\odot}$, with an inert core, i.e., a ball having constant shape, density, pressure, and temperature.  We do include the gravity of the inert core in the entire grid. We performed a test simulation with this RSG setting and without any jets. The outer envelope had some minor fluctuations but no material left the computational grid for the duration of our simulation.  

We assume that a NS, either a single NS or a NS in a tight binary system (Fig. \ref{fig:Schamnatic}), orbits inside the envelope of the star, accretes mass and launches jets.
The orbit is assumed to be circular with a constant radius of $a=  700 R_\odot=4.9 \times 10^{13} \cm = 0.79 R_{\rm RSG}$, neglecting to account for the spiralling-in effect as our goal is to explore other effects, mainly deposition of angular momentum by jets.
We also neglect the gravity of the NS and the variation in the gravity of the envelope due to the envelope's deformation, and simply include a gravitational field due to the initial RSG star which remains constant in time (we do include the gravity of the inner inert core as part of the RSG star). 

We would like to simulate the effects of the jets that the orbiting NS launches, but our limited computational resources do not allow us to resolve the vicinity of the NS, the accretion of mass by the NS, and the launching of jets.
We therefore employ a sub-grid model to inject the jets into the RSG envelope.
We found that our previous sub-grid jet model (e.g., \citealt{Hilleletal2022FB}) is inappropriate for the current study due to its non-conservation of angular momentum. We closely followed the angular momentum evolution of the gas in the numerical grid and found that the numerical scheme for the insertion of jets may significantly affect its value. We have therefore developed and used a new jet-launching sub-grid jet model as we describe next.

In the new scheme both the energy deposition rate and the momentum deposition rate are used as parameters, which is equivalent to determining both the mass deposition rate and the velocity of the jets.
However, in our scheme we neither increase nor decrease the mass of any computational cell; we do not add jet material mass, but rather only change the velocity and internal energy of existing envelope material.
The sum of the magnitudes of the momentum deposition rates of the two jets along the symmetry axis of the jets is $\dot{P}_{\rm 2j} \equiv \vert \dot{P}_{\rm 1} \vert + \vert \dot{P}_{\rm 2} \vert= 1.26 \times 10^{32} \g \cm \s^{-2}$ and the energy deposition rate is $\dot{E}_{\rm 2j} = 3.16 \times 10^{41} \erg \s^{-1}$.

These values result from the following considerations. At the orbit of the NS, $a=700 R_\odot= 4.9 \times 10^{13} \cm$, the initial density of the envelope is $\rho_0(700)= 7.9 \times 10^{-9} \g \cm^{-3}$ and the velocity of the NS is $v_{\rm NS}=59 \km \s^{-1}$, and so the Bondi-Hoyle-Lyttleton (BHL) mass accretion rate is $\dot M_{\rm BHL,0} (700) = 0.27 M_\odot \yr^{-1}$. Following our earlier study \citep{Hilleletal2022FB} we assume that the power of the two jets combined is a fraction $\zeta \ll 1$ of the total accretion power. Namely, 
\begin{equation}
\dot E_{\rm 2j} = \zeta \frac{G M_{\rm NS}}{R_{\rm NS}}\dot M_{\rm BHL,0}. 
\label{eq:JetsPower}    
\end{equation}
Here we take $\zeta=1.2 \times 10^{-4}$, but note that \cite{Hilleletal2022FB} consider values of 
$\zeta \simeq 0.002 - 0.005$ to be more appropriate than the lower value we use here for numerical reasons. For example, if the NS launches the jets at $v_{\rm j}=5 \times 10^4 \km \s^{-1}$ then the mass outflow rate in the two jets for $\zeta=1.2 \times 10^{-4}$ is $\dot M_{\rm 2j}=4 \times 10^{-4} M_\odot \yr^{-1}=1.5 \times 10^{-3} \dot M_{\rm BHL,0} (700)$. For these values a simple calculation gives for the sum of the magnitudes of the two jets' momentum deposition rates $1.26 \times 10^{32} \g \cm \s^{-2}$. 

 We note that the properties of our companion, a NS in these simulations, enter through equation (\ref{eq:JetsPower}). These enter through the ratio of mass to radius and via the efficiency parameter $\zeta$. A black hole will have very similar parameters. A white dwarf would have a mass over radius ratio that is smaller by more than two orders of magnitude and a much lower efficiency $\zeta$ because a white dwarf cannot accrete much mass. A main sequence companion would have a typical mass to radius ratio that is smaller by about five orders of magnitude than the ratio for a NS that we use here. 

We insert the jets inside a cylindrical region centred at the location of the NS and oriented such that its axis is along the axis of the two opposite jets. The cylinder's base is of radius $4 \times 10^{12} \cm$ and its height is $7 \times 10^{12} \cm = 0.14 a$, where $a=700 =  R_\odot4.9 \times 10^{13} \cm$ is the orbital radius of the NS inside the RSG envelope.
The momentum we deposit in the whole volume of the cylinder is in the direction of the axis of the cylinder and away from the NS (opposite directions in the two sides of the equatorial plane). 

In each cycle and for each computational cell inside the jet-injection cylinder we first change the velocity due to the momentum that we add.
The momentum we add in each cycle is $\Delta p_{\rm c}= f_{\rm V,c} \dot P_{\rm 2j} \Delta t$, where $f_{\rm V,c}$ the fraction of the volume which the cell occupies in the cylindrical jet injection region, $\dot P_{\rm 2j}$ was defined above, and $\Delta t$ is the timestep. 
The new velocity is then computed due to the effect of this addition of momentum.
Secondly, we compute the new total energy, i.e., kinetic energy plus internal energy, neglecting gravitational energy which is irrelevant here since it is not changed during the jet insertion in each cell.
The new total energy is $E_{\rm new,c}=E_{\rm old,c}+f_{\rm V,c} \dot E_{\rm 2j} \Delta t$ , where $E_{\rm old,c}$ is the old total energy in the cell. 
Note that the new kinetic energy in the cell is already known from the new velocity computed above.
The third and final step is setting the new internal energy to satisfy energy conservation, i.e., the new internal energy is determined by requiring that the new total energy in the cell be $E_{\rm new,c}$. 
Due to numerical errors, the new internal energy might turn out to be negative in specific grid cells, and we fix these rare cases by setting the new internal energy to some small positive value; in our case we take $10^{-3}$ of the cell energy addition.
 Negative internal energy occurred only in a very small number of cases and the negative internal energy was close to zero. Therefore, this numerical fixing procedure does not affect the global flow and results. 
Except for these rare case, this scheme conserves mass, momentum, and energy.

Our simulations are performed on a cubic Cartesian computational grid with a side of $L_{\rm G} = 5 \times 10^{14} \cm$.  We set outflow conditions on all boundary surfaces of the 3D grid. 
Adaptive mesh refinement (AMR) is employed with a refinement criterion of a modified L\"ohner error estimator (with default parameters) on the $z$-component of the velocity.
The gas in the whole computational domain is an ideal gas with an adiabatic index of $\gamma=5/3$ including radiation pressure.
The centre of the RSG is fixed at the origin.
The smallest cell size in the simulation is 
$L_{\rm G}/128 = 3.90625 \times 10^{12} \cm$.

In sections \ref{sec:Perp} and \ref{sec:Tilted} we will present calculations of angular momentum in the inflated envelope of the RSG. These calculations are performed on the material in the spherical shell between $r_{\rm inert} = 0.2 R_{\rm RSG} = 1.23 \times 10^{13} \cm$ and $r = 1.25 \times 10^{14} \cm$, which is a quarter of the total grid size.  Namely, in calculating the total angular momentum of the envelope we sum the quantity $m_i \vec{r_i} \times \vec{v_i}$ over all cells $i$ with a radius (relative to the center of the RSG) of $r_{\rm inert} < r_i < 1.25 \times 10^{14} \cm$. Here $m_i$, $\vec{r_i}$ and $\vec{v_i}$ are the mass, the location with respect to the center of the RSG, and the velocity of cell $i$. We start at $t=0$ with no envelope rotation, i.e., $\vec{J}(0)=0$.  

\section{The flow structure} 
\label{sec:FlowStructure}
\subsection{The flow structure of perpendicular jets} 
\label{subsec:FlowStructurePerpendicular}

Some prominent flow properties are similar to those in some earlier simulations of jets in CEE (.e.g, \citealt{Shiberetal2019, Hilleletal2022FB, LopezCamaraetal2022MS}). Each of the two opposite jets inflated a hot-low-density bubble. The jet-inflated bubbles expand the envelope in a highly non-spherical manner and eject mass. 

\begin{figure*} [htb!]
\centering
\includegraphics[width=0.44\textwidth]{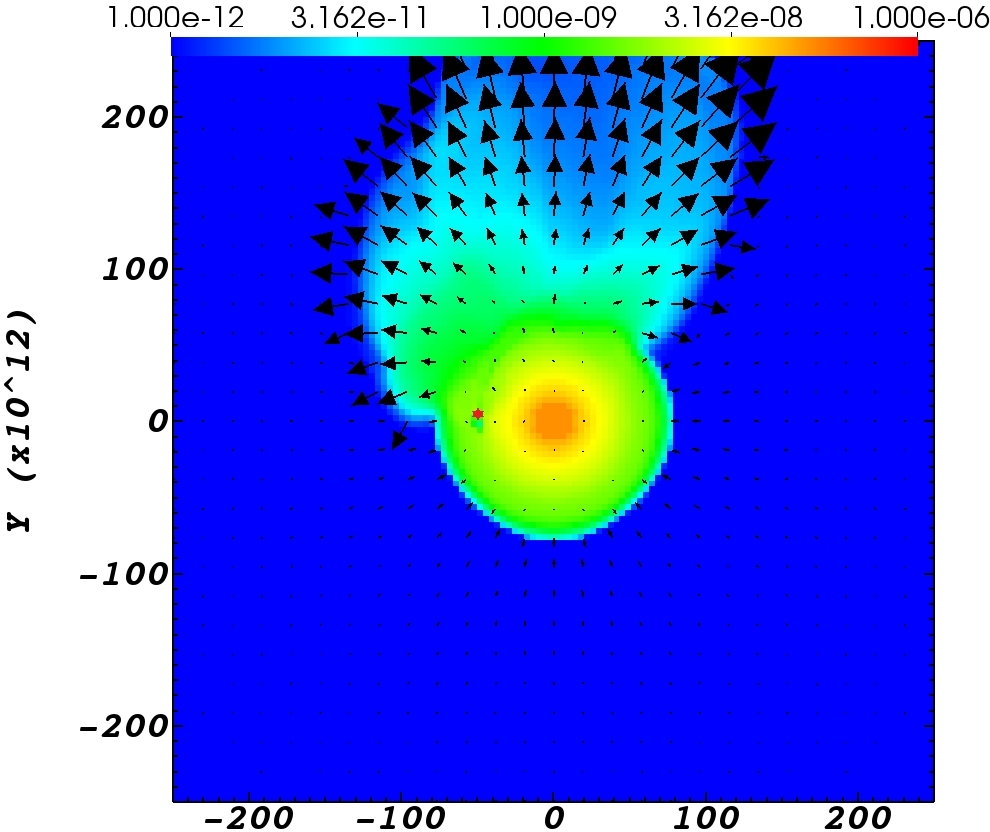}
\includegraphics[width=0.44\textwidth]{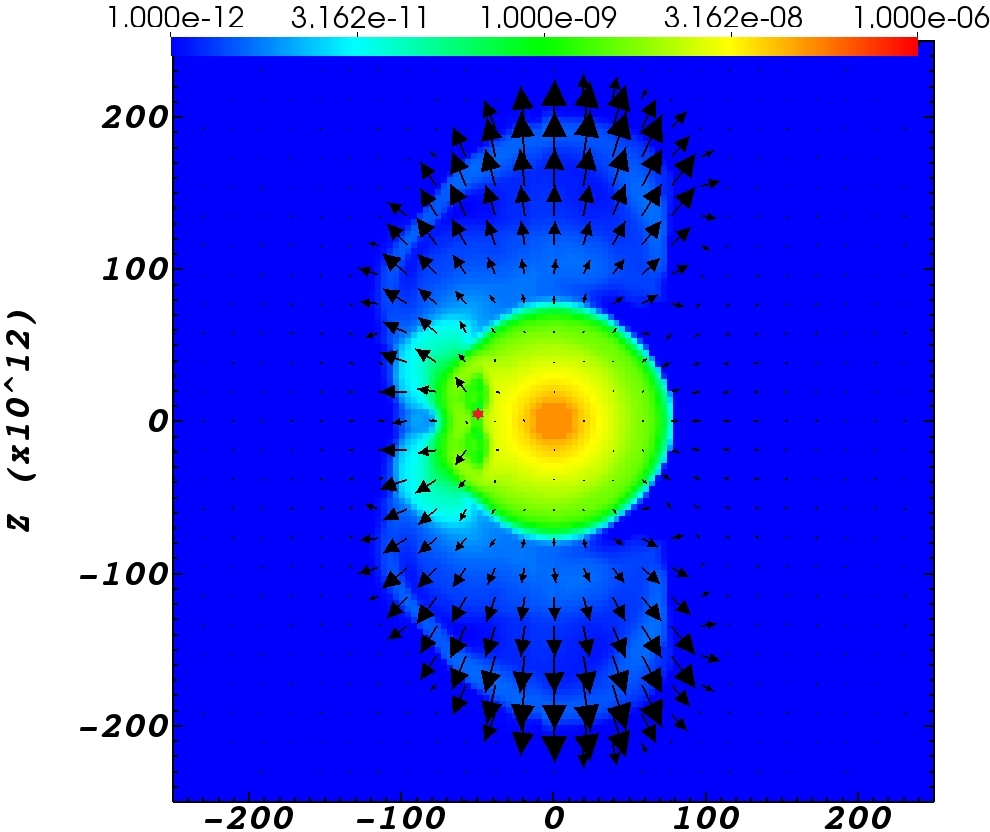}
\includegraphics[width=0.44\textwidth]{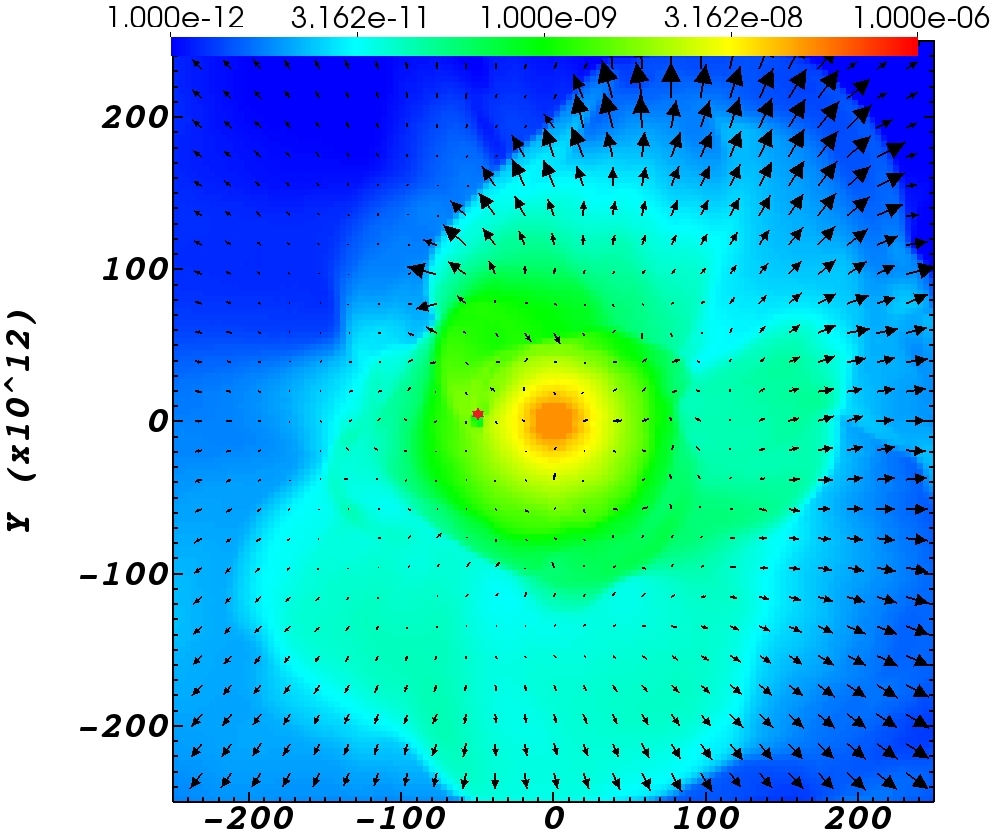}
\includegraphics[width=0.44\textwidth]{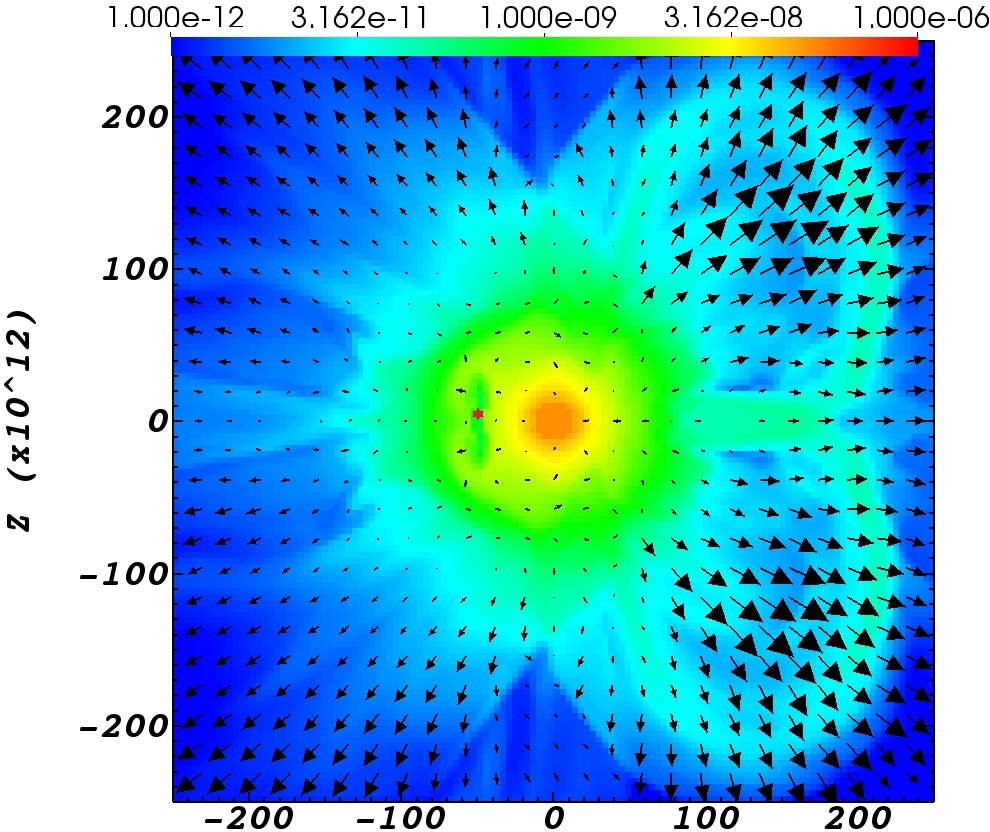}
\includegraphics[width=0.44\textwidth]{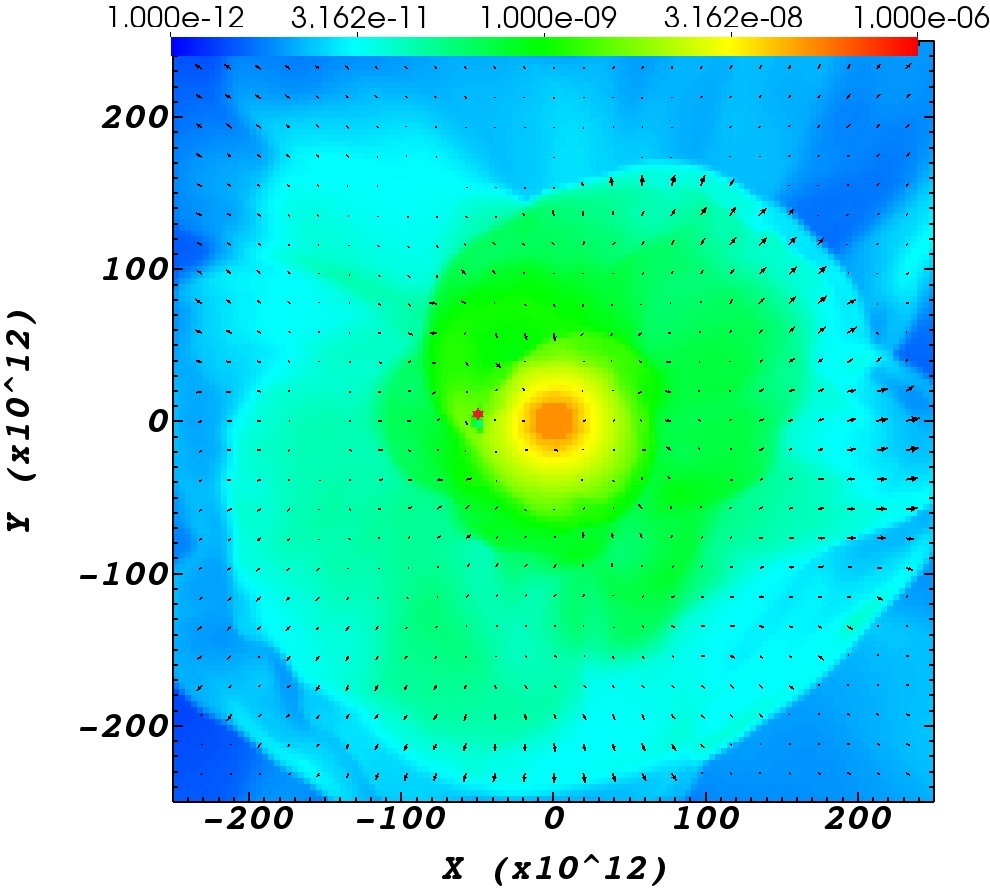}
\includegraphics[width=0.44\textwidth]{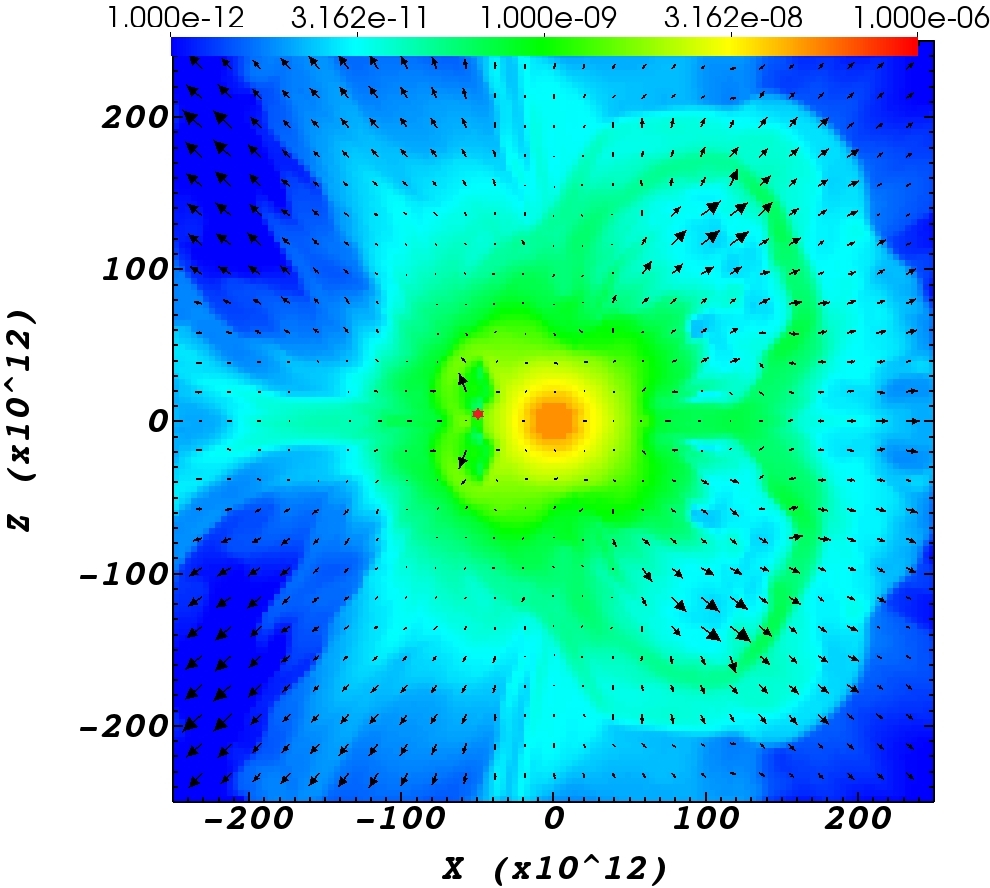}
\caption{Density maps and velocity arrows at three times when the NS completed one quarter of a full orbit (top panels), $5/4$ of an orbit, (middle panels), and $9/4$ of an orbit (bottom panels), for our simulation with jets that are perpendicular to the equatorial plane. Namely, one jet is launched in the direction $(x,y,z)=(0,0,1)$ and the opposite jet is launched in the direction $(x,y,z)=(0,0,-1)$. 
Units on the axes are in $\cm$. One orbit lasts for $1.78 \yr$. The initial location of the NS is at $(x,y,z)=(0, 49 \times 10^{12} \cm, 0)$ and the NS location at the times of the panels is at $(x,y,z)=(-49 \times 10^{12} \cm, 0, 0)$, as marked by the red dot.  The center of the RSG is at $(x,y,z)=(0, 0, 0)$.  The length of each arrow is proportional to the velocity at the respective numerical cell, with minimum length (zero) representing any velocity $\le 10 \km \s^{-1}=10^6 \cm \s^{-1}$ in all panels, and maximum length that is in the upper-left panel represents a velocity value of $180 \km \s^{-1}$. 
The left-column panels show the density in the plane $z= 3 \times 10^{12} \cm$ and the right-column panels show the density in the plane $y=0$. 
The density colour coding is according to the upper colour bar in units of $\g \cm^{-3}$ and in the range of $10^{-12} \g \cm^{-3}$ (deep blue) to $10^{-6} \g \cm^{-3}$ (deep red).
}
\label{fig:Dens_no_tilt}
\end{figure*}

In Fig. \ref{fig:Dens_no_tilt} we present the density and velocity maps of the simulation where we launch the jets perpendicular to the orbital plane. 
The left column presents the maps at three times in the plane $z=3 \times 10^{12} \cm$ that is parallel to the equatorial plane, while the right column presents the density maps in the meridional plane $y=0$ at the same three times. 
The three times are at quarter of an orbit, $t=0.25P_{\rm orb}=1.78 \yr /4$ after the starting time of our simulation $t=0$, at $t=1.25P_{\rm orb}$ and at $t=2.25P_{\rm orb}$. 
The location of the NS that launches the jets at these three times is at $(x,y,z)=(-49 \times 10^{12} \cm, 0, 0)$, i.e., left to the center of each of the six panels,  as marked by the red dot.  
On the left panels we can see the bubble by a small green area surrounded by a thin yellow region (higher density than inside the bubbles) left to the large yellow ring. 

We can better see the bubbles in the three right panels. Left to the center of each of the right panels we can see the bubbles above and below the NS. The length of each bubble (in the $z$-direction) is $\simeq 30 \times 10^{12} \cm$. To the right of the center of the grid (the center of the RSG) and in the lower two panels we can see the cross section of the bubbles on the meridional plane half an orbit after the NS has passed there. The bubbles are much larger now because they expand. 

There is an interesting effect with the cross section of the bubbles half an orbit after the passages of the NS. Consider the bubbles to the right in the middle-right panel of Fig. \ref{fig:Dens_no_tilt}. The panel is at $t=1.25 P_{\rm orb}$, but the jets inflated the bubbles on the right about an half orbit earlier, at $t=0.75 P_{\rm orb}$, when the NS was in that region, i.e., at $(x,y,z)=(-49 \times 10^{12} \cm, 0, 0)$. Half an orbit after the first passage of the NS through $(x,y,z)=(-49 \times 10^{12} \cm, 0, 0)$ (right to the center) the bubbles are very large, as we see in the middle-right panel of Fig. \ref{fig:Dens_no_tilt}; the bubbles almost touch the edge of the numerical grid. However, at $t=2.25 P_{\rm orb}$, namely half an orbit after the second passage of the NS through that point (which took place at $t=1.75 P_{\rm orb}$) the bubbles are smaller (right to the center of the lower-right panel of Fig. \ref{fig:Dens_no_tilt}) than half an orbit after the first passage (middle-right panel of Fig. \ref{fig:Dens_no_tilt}). This is because in the second passage of the NS at that location, the jets have a larger envelope to interact with. Namely, in the first NS passage the jets that the NS launched inflated the envelope, and in the second passage the jets have more envelope mass that they interact with relative to the first passage.  Note that this is the case because most of the volume of the inflated bubbles on the right in the middle-right panel of Fig. \ref{fig:Dens_no_tilt} is outside the original RSG envelope. When this happens in the first time there is no gas outside the original RSG radius and the inflation of the bubbles is into practically an empty space. In the second passage there is gas outside the original RSG envelope that slows down the inflation of the bubbles. 

Unfortunately, this complicated jet-envelope interaction also implies a much longer computational time that prevents us from presenting later times with the computational resources we have. 

\subsection{The flow structure of tilted jets} 
\label{subsec:FlowStructureTilted}
The flow structure of the tilted jets has a much larger departure from a spherical morphology, in particular in losing the mirror symmetry about the equatorial plane. This is expected as in this simulation the launching of the jets does not have a mirror symmetry about the equatorial plane. We assume that the jets' axis has a constant direction that is maintained by a tight (inner) binary system (Fig. \ref{fig:Schamnatic}). Namely, the NS is a member of a tight binary system that enters the envelope of the RSG star (a triple star system;  \citealt{Soker2022misalignment}). We launch both jets from the location of the NS that orbits the center of the RSG star, one jet in the direction of  $(x,y,z)=(-1/\sqrt{2},0,1/\sqrt{2})$ and the opposite jet in the direction of $(x,y,z)=(1/\sqrt{2},0,-1/\sqrt{2})$. Namely, while the NS is on the half space of $x<0$ the jet that is above the equatorial plane ($z>0$) is launched towards the outer parts of the RSG envelope, while when the NS is in the half space $x>0$ the jet that is launched below the equatorial plane ($z<0$) is launched outwards in the RSG envelope. 
 The highly distorted envelope requires a longer computational time. The simulation of the tilted jets took about ten weeks on our cluster of 280 CPU cores. 
 
\begin{figure*} [htb!]
\centering
\includegraphics[width=0.44\textwidth]{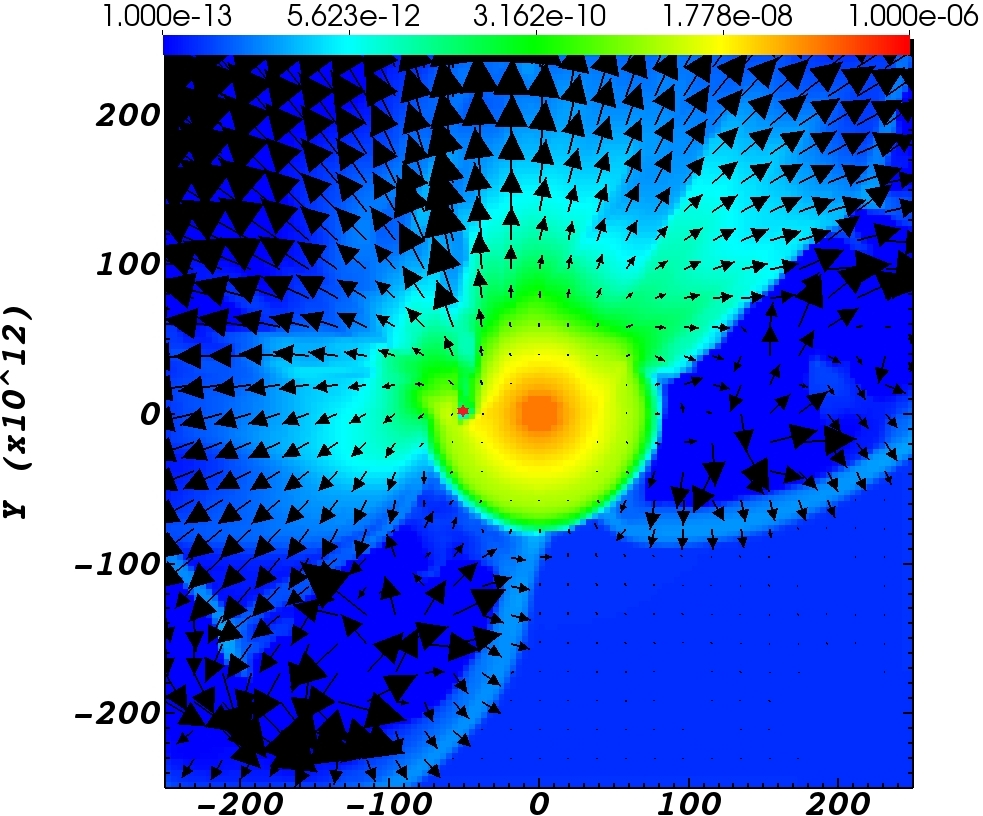}
\includegraphics[width=0.44\textwidth]{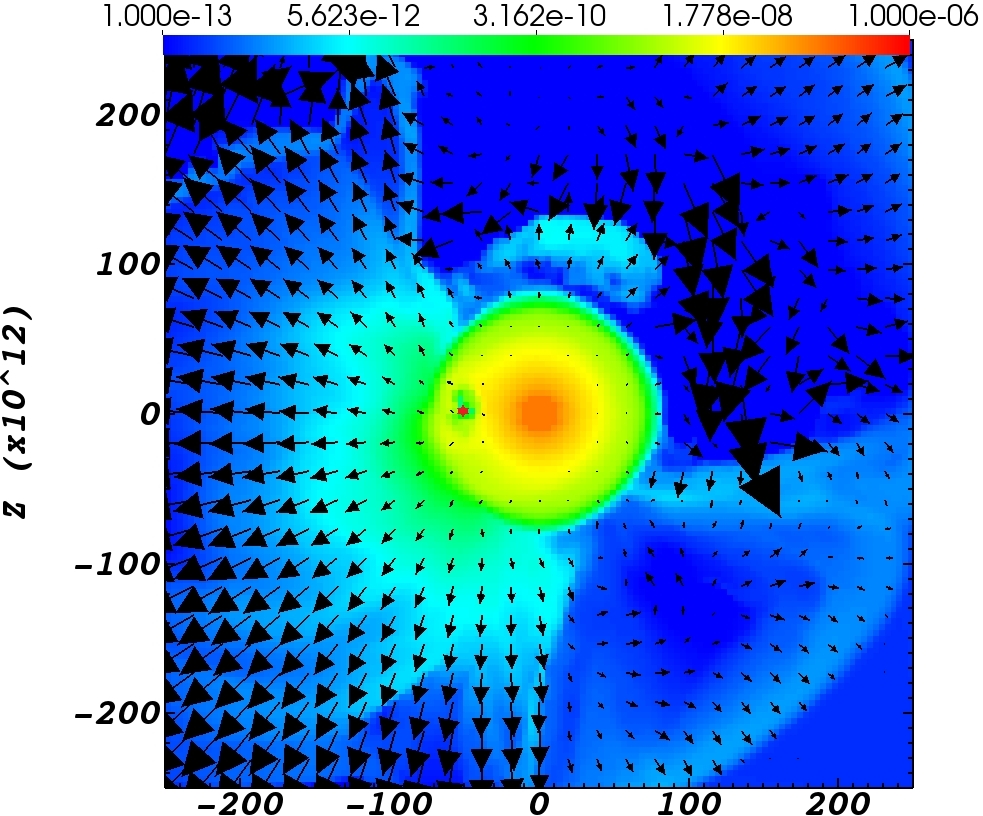}
\includegraphics[width=0.44\textwidth]{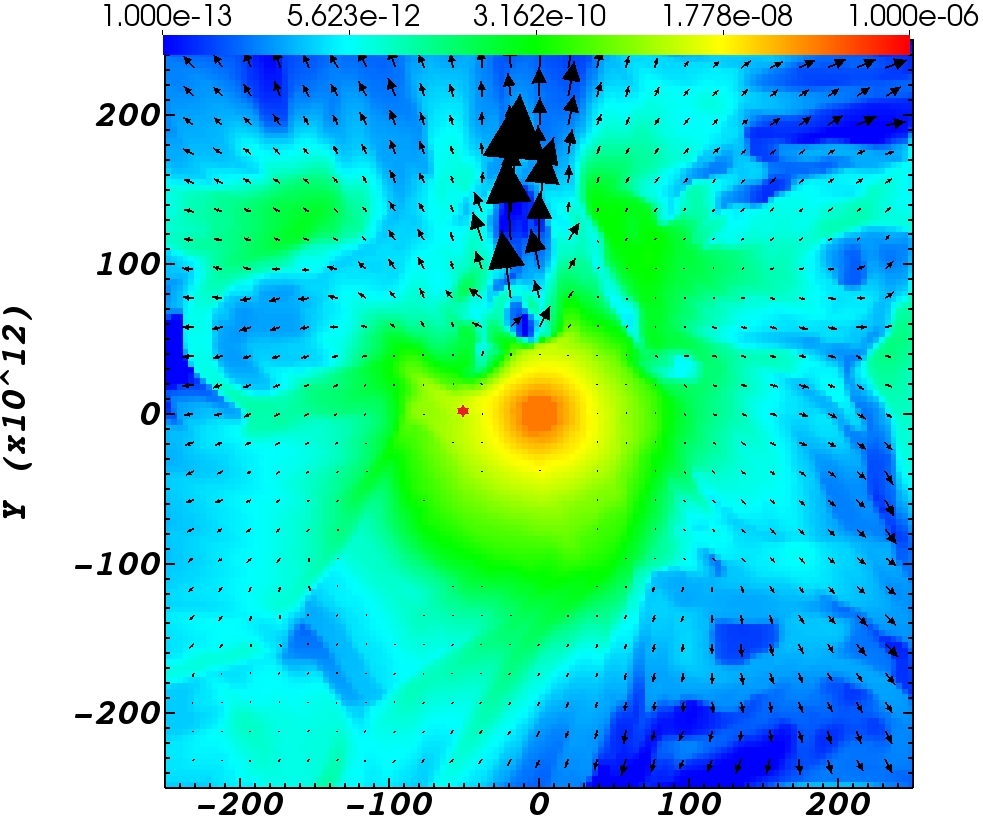}
\includegraphics[width=0.44\textwidth]{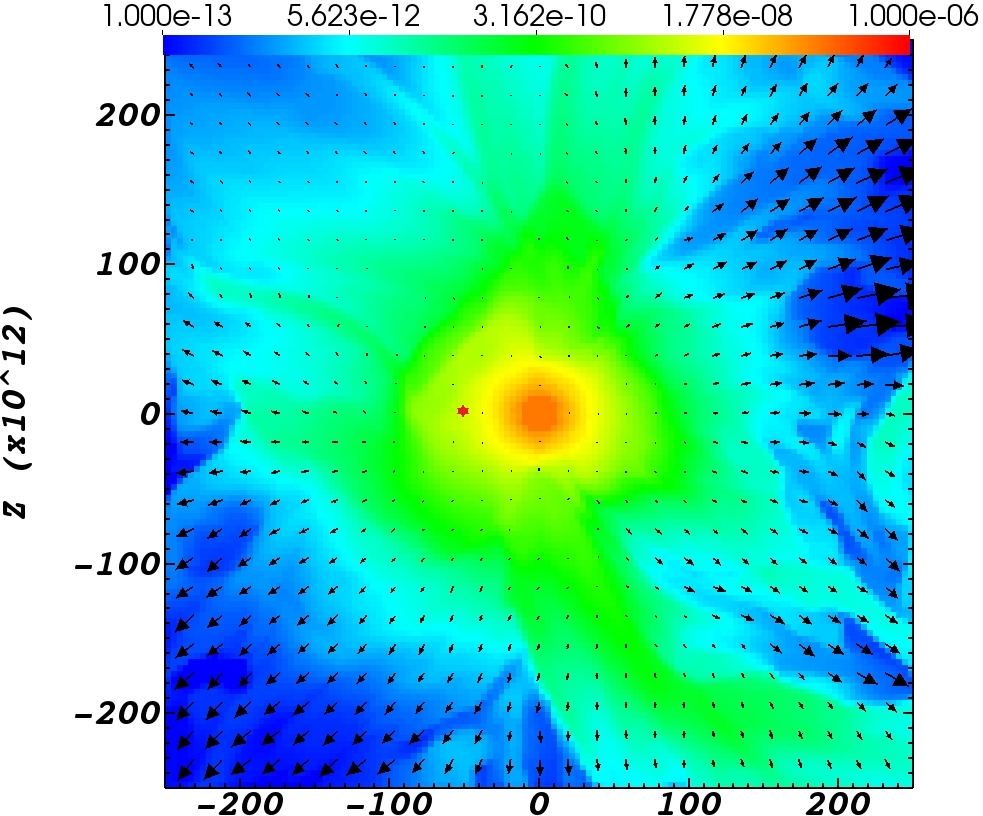}
\includegraphics[width=0.44\textwidth]{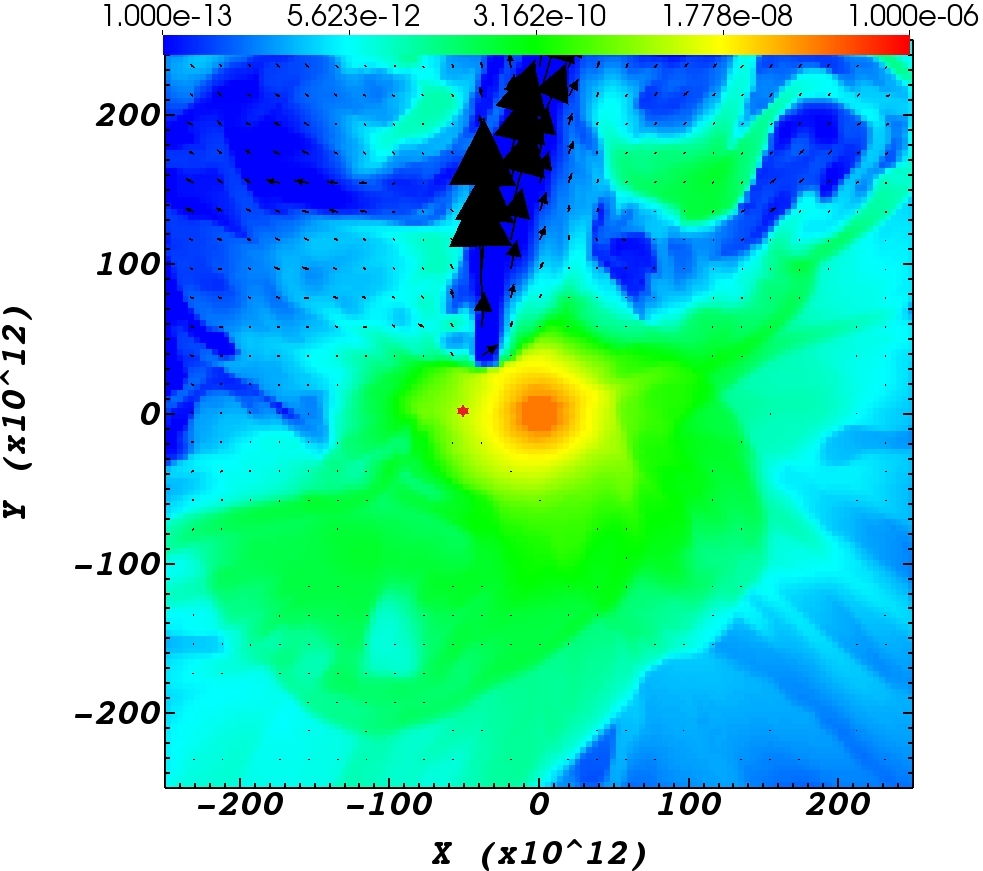}
\includegraphics[width=0.44\textwidth]{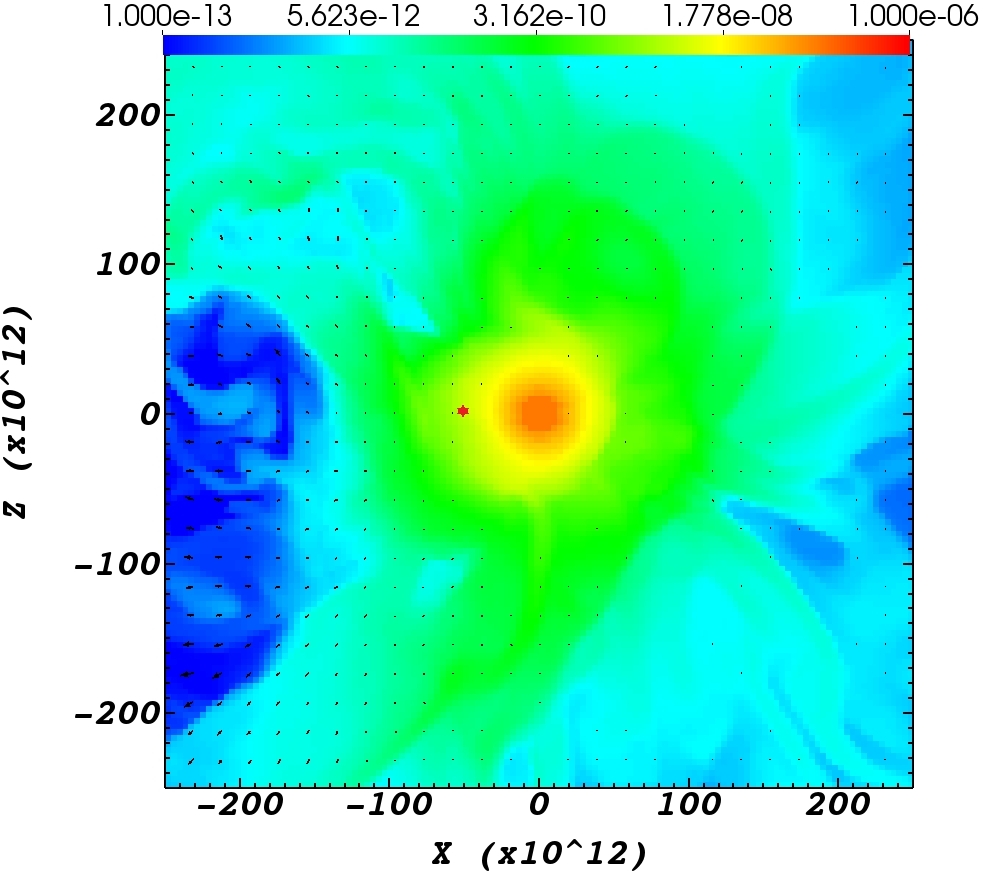}
\caption{Similar to Fig. \ref{fig:Dens_no_tilt} but for the case where the jets are launched with a tilt angle of $45^\circ$ with respect to the orbital plane and in a plane parallel to the $(x,z)$ plane. Namely, one jet is launched in the direction of $(x,y,z)=(-1/\sqrt{2},0,1/\sqrt{2})$ and the opposite jet is launched in the direction $(x,y,z)=(1/\sqrt{2},0,-1/\sqrt{2})$. 
 Note also that the blue color represents all densities of $\rho \le 10^{-13} \g \cm^{-3}$ rather than  $\rho \le 10^{-12} \g \cm^{-3}$ as in the other figures. The scale of arrow-lengths is the same as in Fig. \ref{fig:Dens_no_tilt}, but here the velocities are higher and there are many arrows with the maximum length that represents any velocity of $\ge 600 \km \s^{-1}=6 \times 10^7 \cm \s^{-1}$.   
}
\label{fig:Dens_45deg_tilt}
\end{figure*}

In Fig. \ref{fig:Dens_45deg_tilt} we present the density and velocity maps of the tilted-jets simulation at the same three times and in the same two planes as in Fig. \ref{fig:Dens_no_tilt} for the perpendicular-jets simulation  (note the lower density scale in Fig. \ref{fig:Dens_45deg_tilt}).  
In the upper panel of Fig. \ref{fig:Dens_45deg_tilt} we see the bubble. However, the jets are titled and the flow differs from that of Fig. \ref{fig:Dens_no_tilt}. Consider the upper-right panel that shows the density map in the meridional plane at $t=0.25 P_{\rm orb}$. At the position of the NS, to the left of the RSG center, one jet is launched to the upper left direction, i.e., towards the outer zones of the envelope, while the opposite jet is directed to the lower right towards the denser zones of the envelope. It seems that the jet that is directed into the RSG envelope strongly interacts with the deeper envelope to form a very high-pressure zone that accelerates envelope gas towards the lower part of the grid below the equatorial plane ($z<0$). 
The highly non-spherical (`messy') envelope structure results from the highly unequal interaction of the two jets with the envelope.

\begin{figure*} [htb!]
\centering
\includegraphics[width=0.44\textwidth]{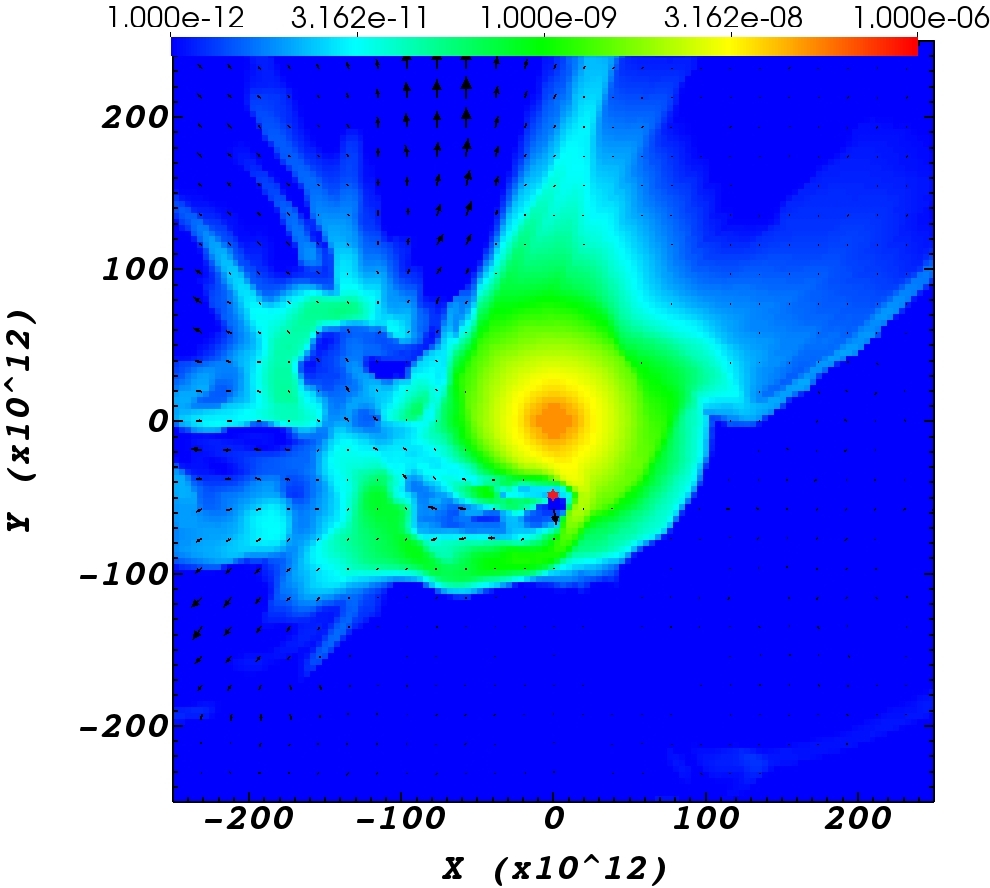}
\includegraphics[width=0.44\textwidth]{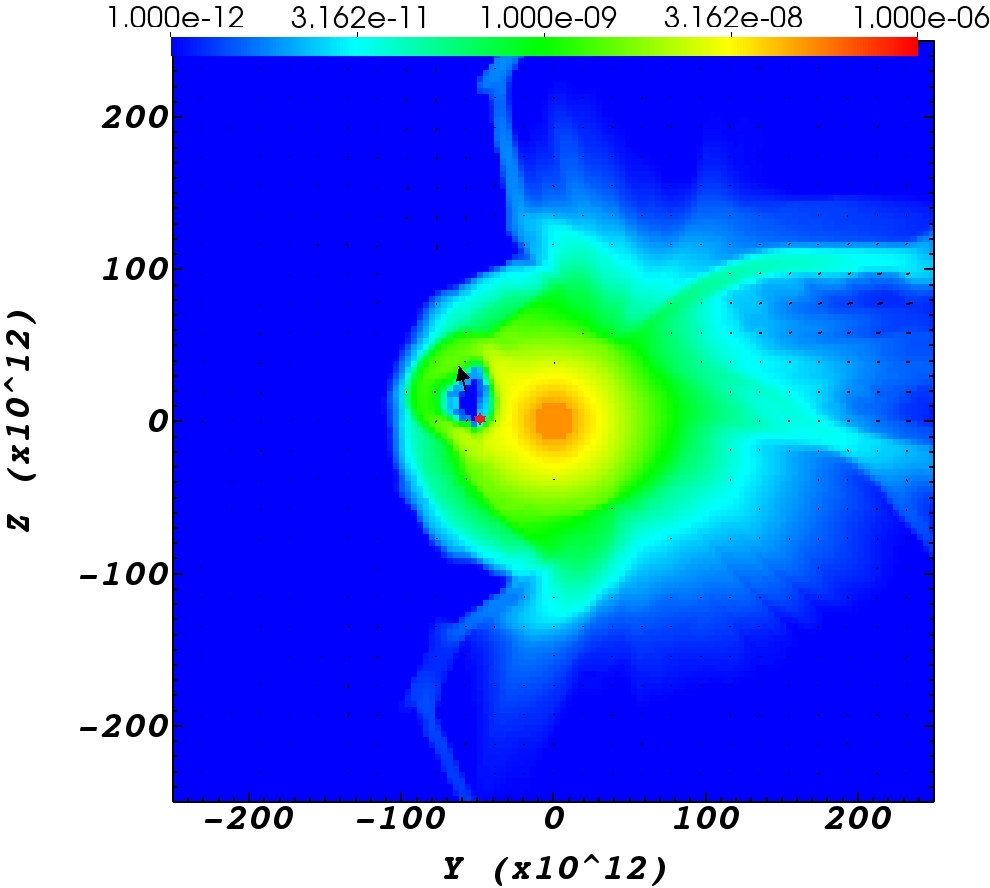}
\includegraphics[width=0.44\textwidth]{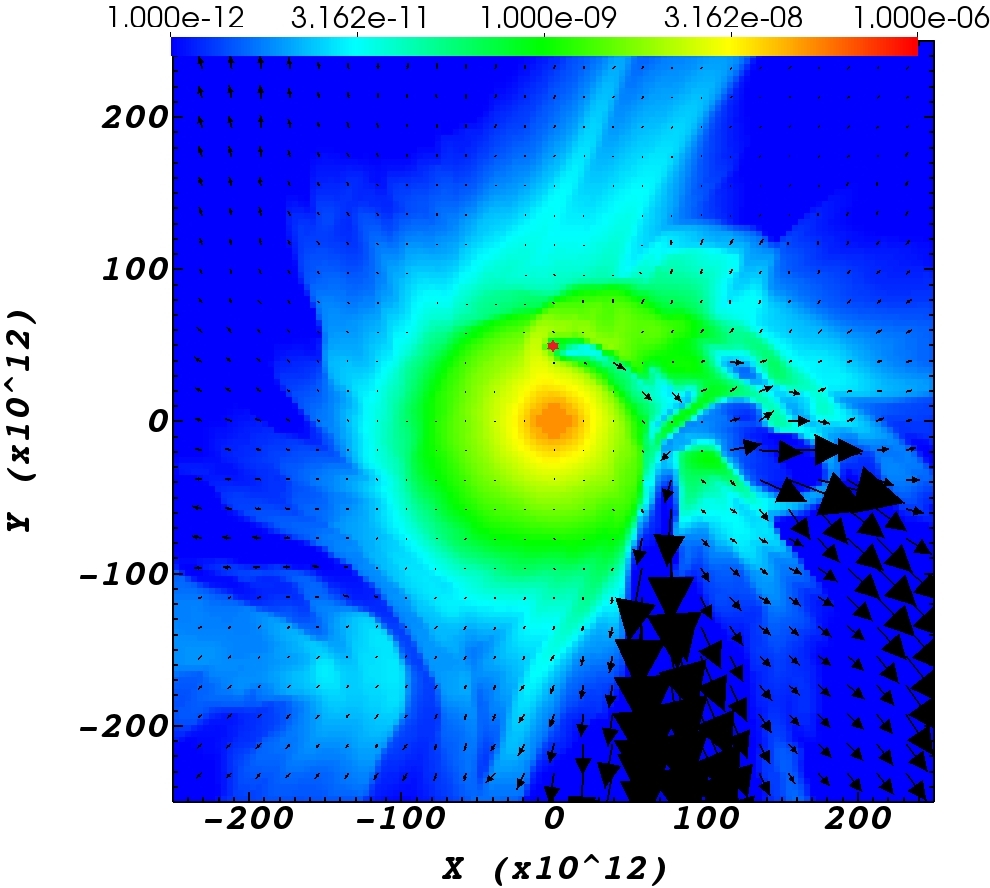}
\includegraphics[width=0.44\textwidth]{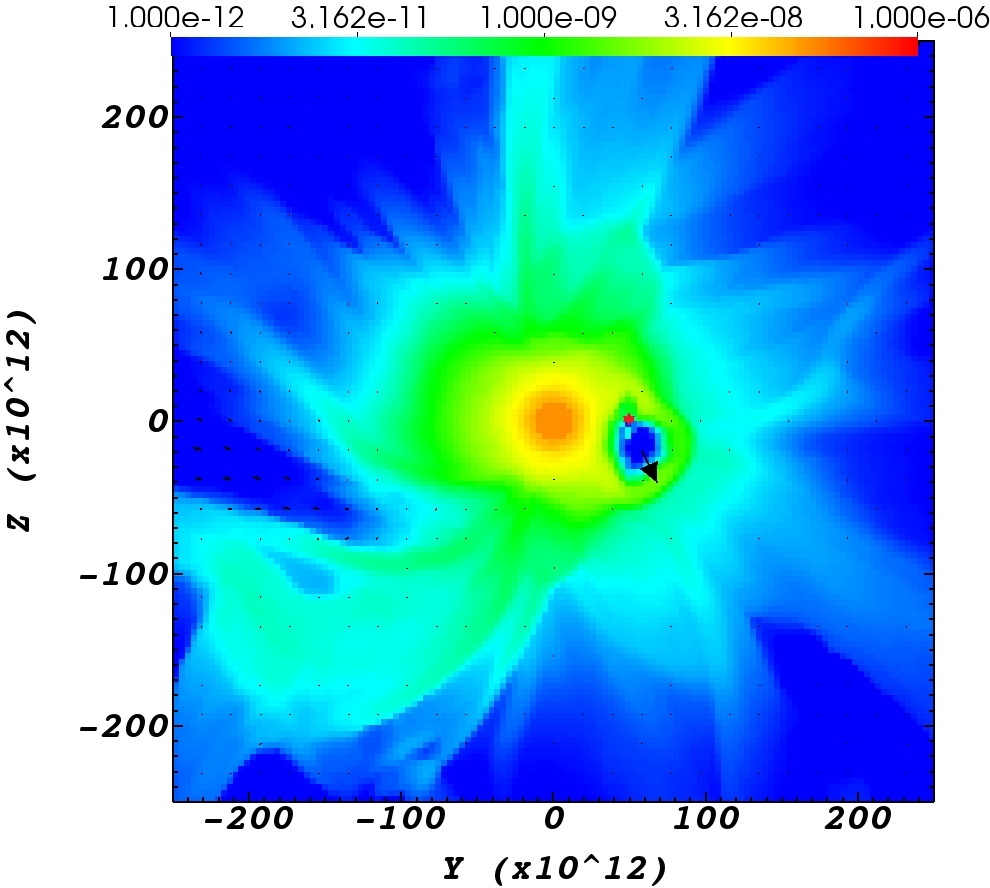}
\caption{Density maps with velocity arrows of the tilted-jets simulation in the plane $z= 6 \times 10^{12} \cm$ (left panels) and in the meridional plane $x=0$ (right panels).  Scale of arrow lengths as in Fig. \ref{fig:Dens_45deg_tilt}.  Upper panels are at $t=0.5P_{\rm orb}$ and the lower panels are at $t=P_{\rm orb}$. The NS locations at the times of the upper and lower panels are $(x,y,z)=(0, -49 \times 10^{12} \cm, 0)$ and $(x,y,z)=(0, 49 \times 10^{12} \cm, 0)$, respectively,  as marked by the red dot in each panel.  The density colour coding is according to the upper colour bar and in units of $\g \cm^{-3}$ from $10^{-12} \g \cm^{-3}$ (deep blue) to $10^{-6} \g \cm^{-3}$ (deep red).
}
\label{fig:rho_v_XY_YZ_tilt}
\end{figure*}

To further demonstrate the complicated flow structure of the tilted-jets simulation, we present in Fig. \ref{fig:rho_v_XY_YZ_tilt} density and velocity maps at $t=0.5 P_{\rm orb}$ and at  $t=P_{\rm orb}$ in the planes $z= 6 \times 10^{12} \cm$ that is parallel to the equatorial plane and in the meridional plane $x=0$.  

Our main goal is to explore the deposition of angular momentum into the envelope by the jets. We therefore do not go further with the analysis of the flow structure here. We will present more velocity maps later where we study the deposition of angular momentum by jets. 

\section{Angular momentum by perpendicular jets}
\label{sec:Perp}

In this section we study the angular momentum that the jets deposit to the RSG envelope for the simulation of perpendicular jets, i.e., we launch the jets perpendicular to the the orbital plane, which means a zero tilt to the orbital angular momentum axis. 

\begin{figure*} 
\centering
\includegraphics[width=0.56\textwidth]{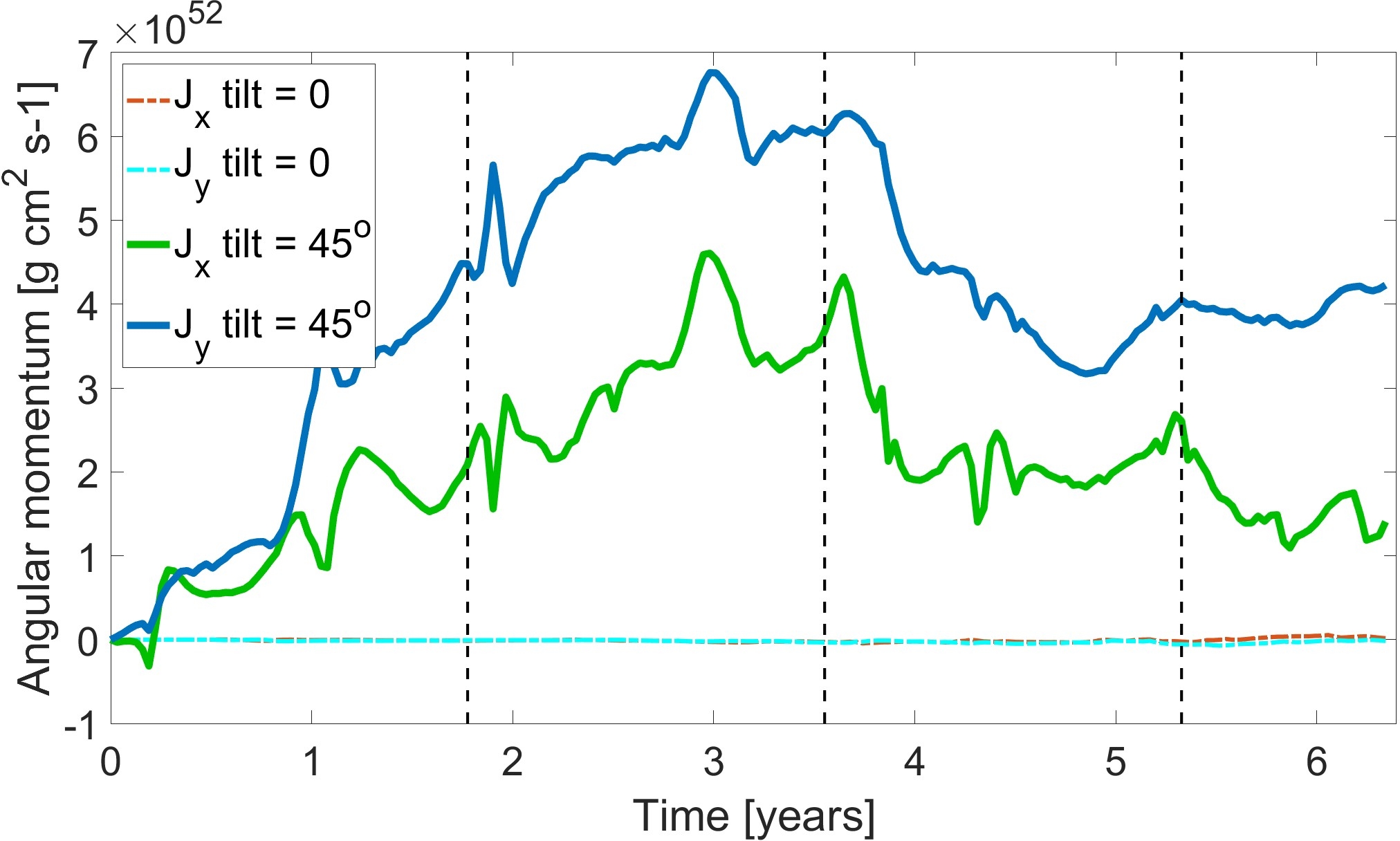}
\includegraphics[width=0.56\textwidth]{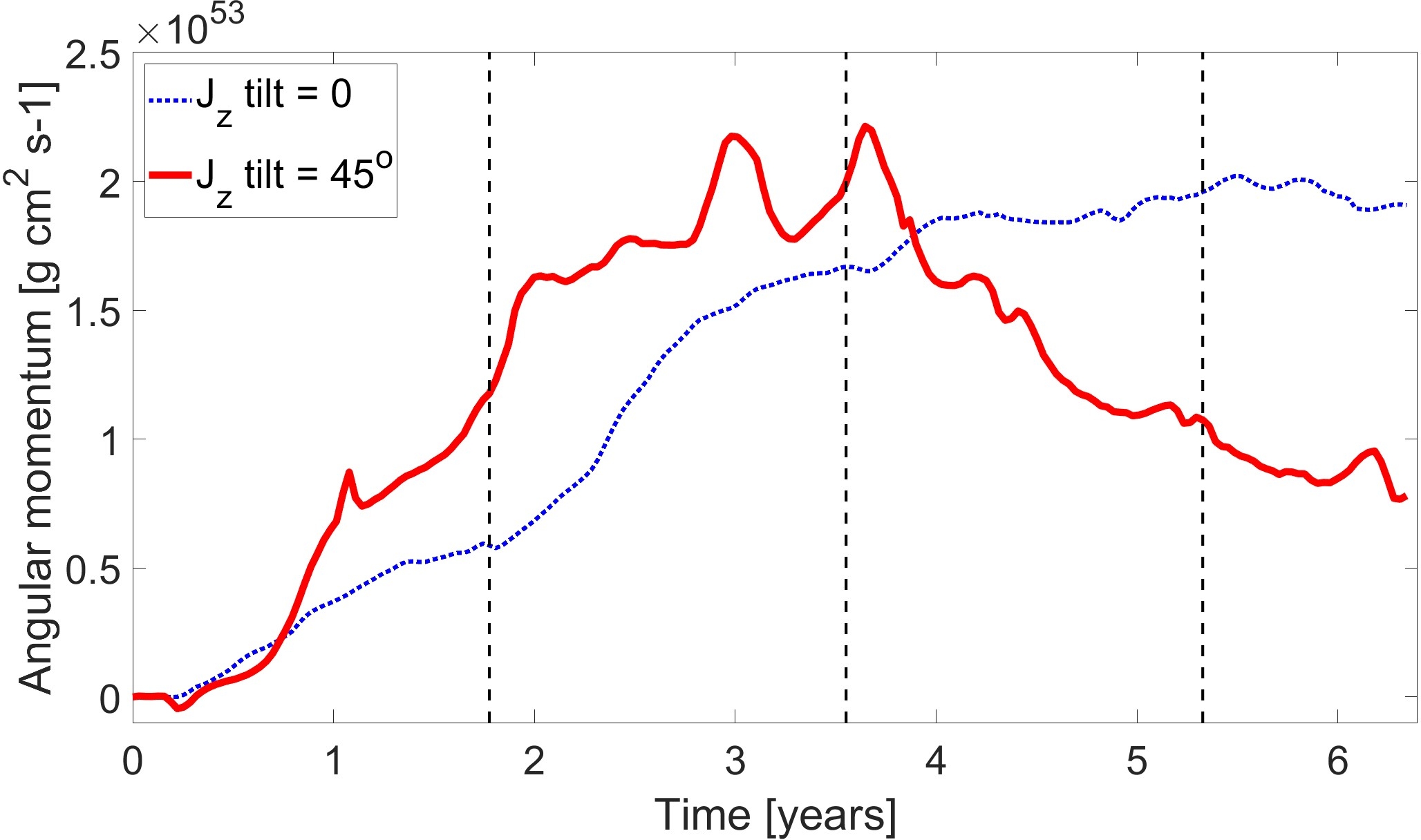}
\includegraphics[width=0.56\textwidth]{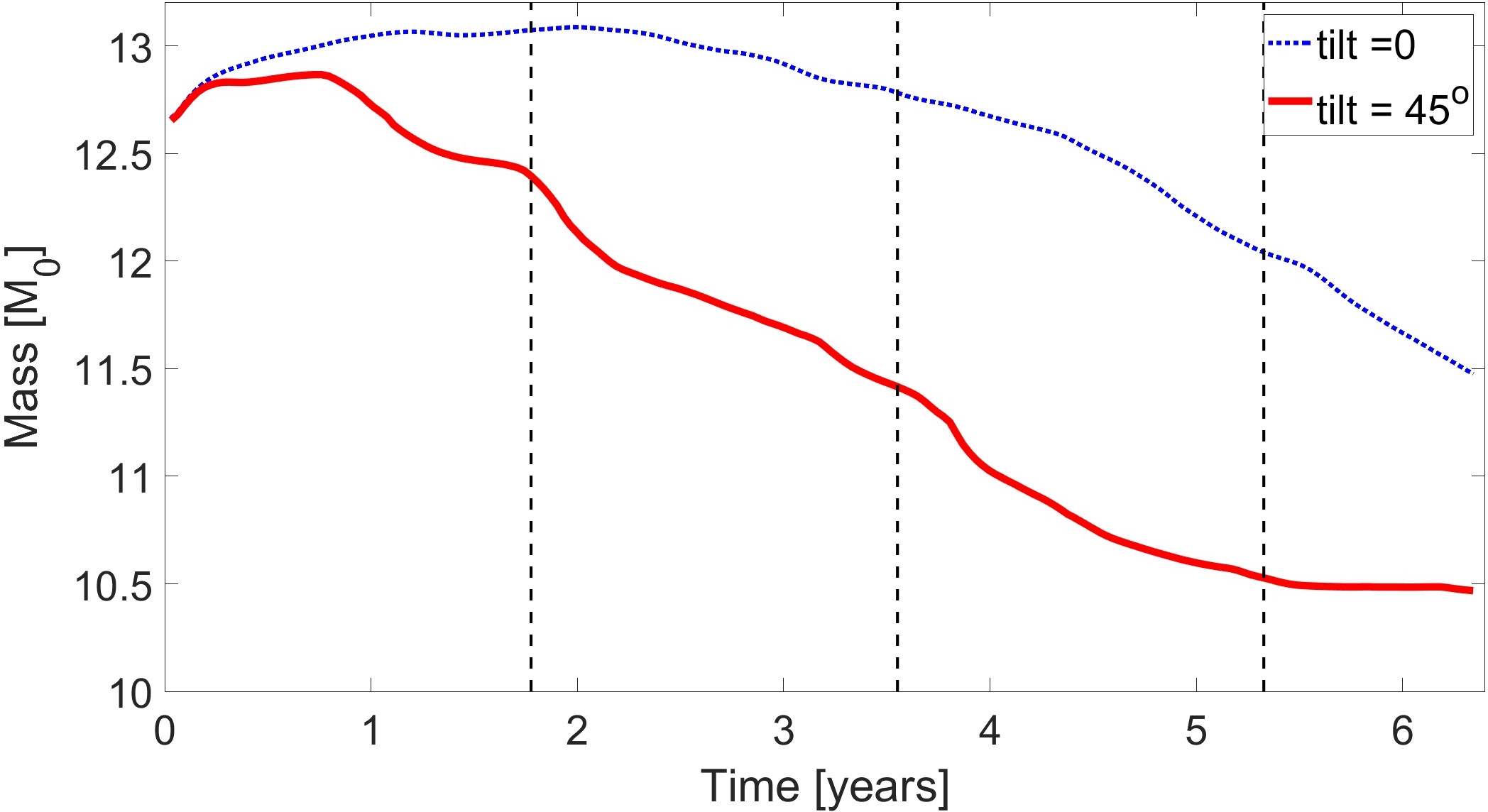}
\caption{ Upper panel:
The $x$ and $y$ components of the angular momentum that the jets deposit to the RSG envelope as function of time. Dotted lines (almost horizontal in the upper panel) are for the simulation where we launch jets perpendicular to the equatorial plane (zero tilt), and solid lines are for the simulation with $45^\circ$ tilted jets. We calculate the angular momentum in the volume between the numerically inert core at $r_{\rm inert} = 0.2 R_{\rm RSG} = 1.23 \times 10^{13} \cm$ and the radius 
$r = 1.25 \times 10^{14} \cm$.
Middle panel: The $z$ component of the angular momentum hat the jets deposit to the envelope. Note that the upper and middle panels have different vertical scales. 
Bottom panel: The total mass in the grid as a function of time for the two simulations. 
The dashed vertical lines denote one, two, and three orbital periods of the NS.  
}
\label{fig:J_components_and_mass}
\end{figure*}

In Fig. \ref{fig:J_components_and_mass} we present the angular momentum components in the inflated envelope (two upper panels), and the mass in the entire grid (lower panel) as function of time.  We take the relevant RSG envelope mass at each time step to be the mass inside  the volume between the inert core and about twice the initial RSG radius, i.e., we consider the inflated envelope to be inside the shell $ 0.2 R_{\rm RSG} = 1.23 \times 10^{13} \cm < r < 2 R_{\rm RSG} = 1.25 \times 10^{14} \cm$.  We calculate the angular momentum of this mass about the center of the RSG star. We do not include the angular momentum of the mass that is outside this volume nor the mass that left the grid. 

From the two upper panels of Fig. \ref{fig:J_components_and_mass} we learn that the perpendicular jets deposit angular momentum to the RSG envelope that is practically parallel to the orbital angular momentum. Namely, the main component of the jet-deposited angular momentum is $J_{\rm z}({\rm per}) >0$, while $\vert J_x({\rm per}) \vert \simeq \vert J_y({\rm per}) \vert \ll J_{\rm z}({\rm per})$. This is expected from symmetry considerations. Moreover, it demonstrates that the numerical scheme conserves angular momentum as the $J_x$ and $J_x$ components stay very small as expected. 

We present the $z$-component (parallel to the orbital angular momentum) of the angular momentum that the jets of the perpendicular simulation deposit to the envelope, $J_z({\rm per})$, by the blue-dashed line in the middle panel of Fig. \ref{fig:J_components_and_mass}. In the first $t \simeq 0.16 \yr$ the angular momentum $J_z({\rm per})$ stays zero. This further demonstrates that our jet-launching numerical scheme conserves angular momentum. At $t \simeq 0.16 \yr$ mass starts to flow out from the sphere $r=1.25 \times 10^{14} \cm$ inside which we sum the envelope angular momentum. At this time the stellar angular momentum starts to increase because the mass that leaves that sphere has a negative angular momentum, $J_z({\rm per,out})<0$. Namely, the mass that leaves the grid because of the jets-envelope interaction carries angular momentum with an opposite sign to that of the orbital angular momentum.

\begin{figure} [htb!]
\centering
\includegraphics[width=0.46\textwidth]{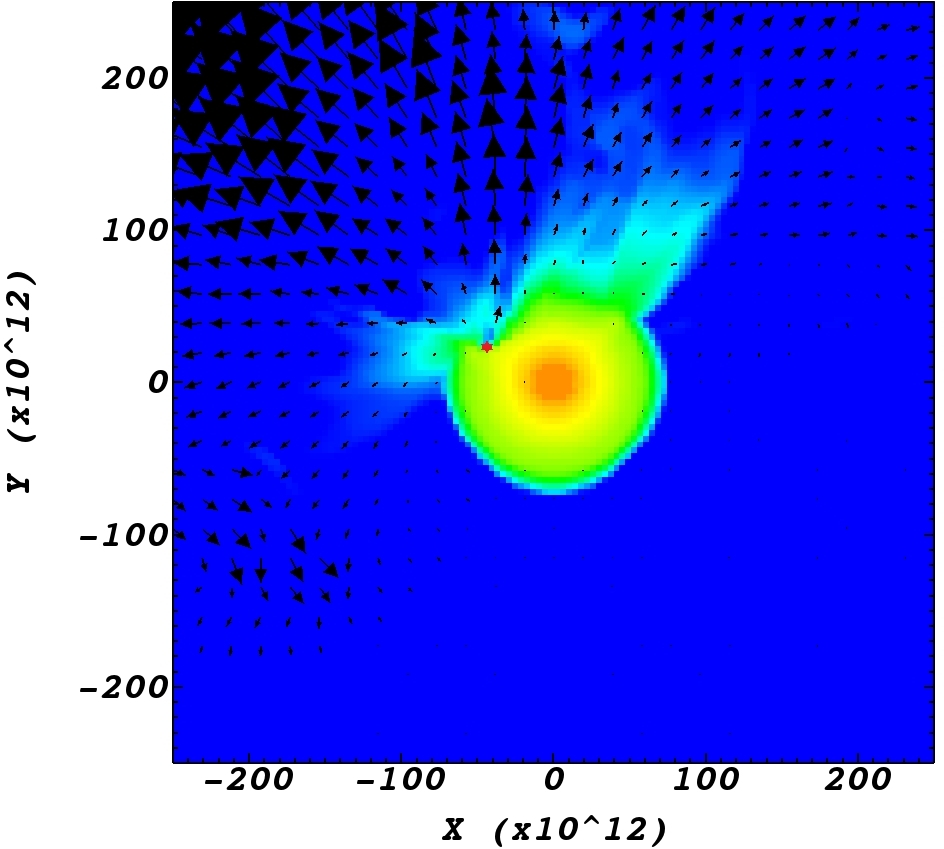}
\caption{Density and velocity map in the plane $z=3 \times 10^{12} \cm$, at $t=0.3 \yr$, demonstrating the ejection of envelope gas opposite to the orbital motion of the NS in the simulation with jets that are perpendicular to the equatorial plane (see also  Fig. \ref{fig:Dens_no_tilt}). The NS started at $t=0$ above the center in the figure at $(x,y,z)=(0, 49 \times 10^{12} \cm, 0)$, and moves to the left. Density color coding is as in Fig. \ref{fig:Dens_no_tilt}.  Arrow lengths are as in Fig. \ref{fig:Dens_45deg_tilt}.  
}
\label{fig:envelope_outflow_perp}
\end{figure}

We demonstrate this negative-angular momentum outflow in Fig. \ref{fig:envelope_outflow_perp} that presents the density and velocity map in the plane $z=3 \times 10^{13} \cm$ (parallel to the equatorial plane) at $t=0.3 \yr=0.17 P_{\rm orb}$. Again, the NS starts at $(x,y,z)=(0,4.9\times 10^{13} \cm, 0)$, and moves in this plane counter-clockwise, i.e., it starts to move to the $x<0$ direction. Despite the counterclockwise orbital motion we see envelope gas ejected towards regions of $x>0$, i.e., clockwise (green zone extending to the upper right). In total, the ejected gas carries negative angular momentum and the jets spins-up the RSG envelope in the same direction as the orbital angular momentum, at least at early times. 

At the end of the simulation the rate of angular momentum deposition almost vanishes $\dot J_z({\rm per}) \simeq 0$. By that time the jets interaction with the envelope has deposited a total of $J_z ({\rm per}) \simeq 2 \times 10^{53} \g \cm^2 \s^{-1}$. We can compare this value to the total angular momentum of the binary system at the beginning of our simulation at $a=700 R_\odot=4.9 \times 10^{13} \cm$, $J_z ({\rm orb,700}) \simeq 7.3 \times 10^{53} \g \cm^2 \s^{-1}$, or at the onset of the CEE at $a=R_{\rm RSG}=881 R_\odot = 6.13 \times 10^{13} \cm$, $J_z ({\rm orb,881}) = 8.5 \times 10^{53} \g \cm^2 \s^{-1}$. Namely, $J_z  ({\rm per}) \simeq 0.25 J_z ({\rm orb})$. 

\begin{figure} [htb!]
\centering
\includegraphics[width=0.46\textwidth]{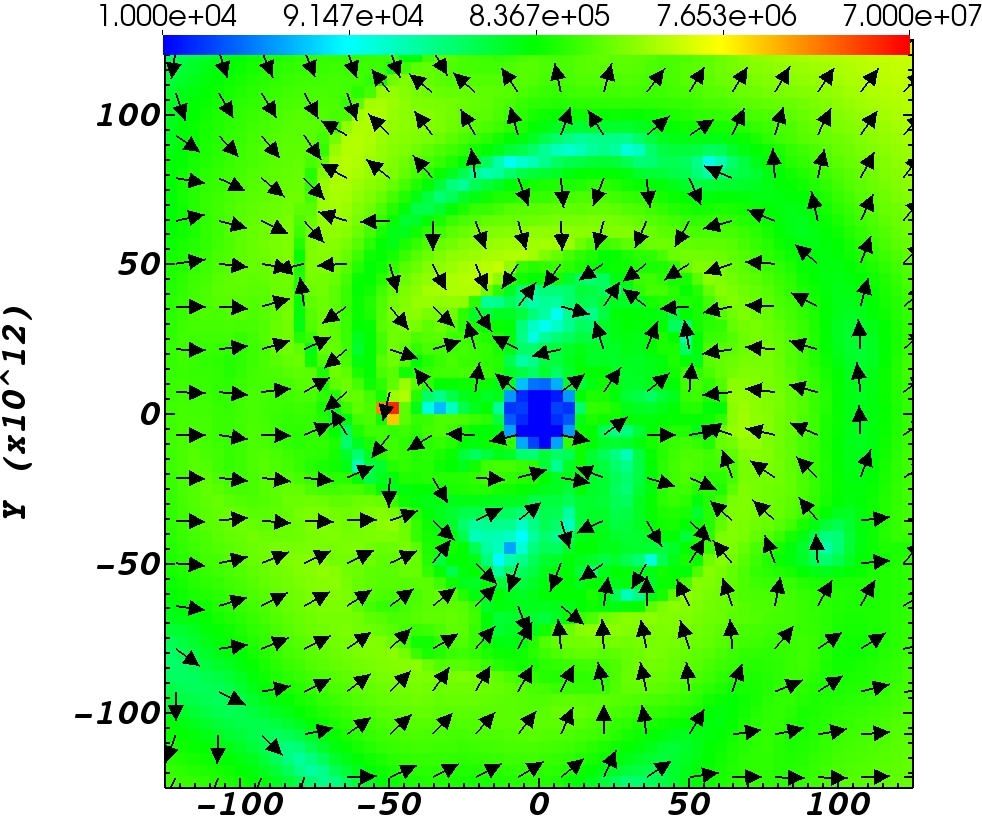}
\includegraphics[width=0.46\textwidth]{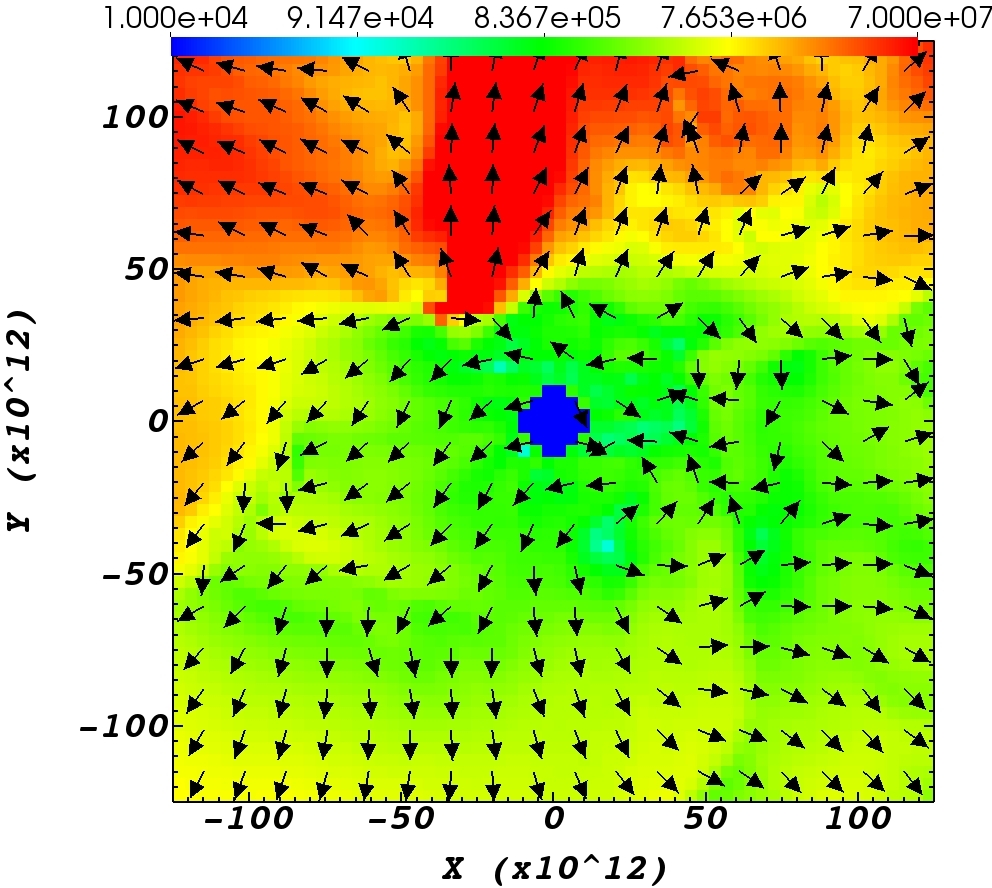}
\caption{Velocity maps at $t=2.25 P_{\rm orb}$ in the plane $z= 3 \times 10^{12} \cm$, for the perpendicular-jet simulation (upper) and the tilted-jet simulation (lower). 
These panels are at the same plane and at the same time as the lower-left panels of Fig. \ref{fig:Dens_no_tilt} and \ref{fig:Dens_45deg_tilt}, respectively, but zooming on the inner region. Velocity magnitudes are according to the color-bar from $0.1 \km \s^{-1}$ (deep blue) to $700 \km \s^{-1}$ (deep red).    
}
\label{fig:EnvelopeRotation}
\end{figure}

In the upper panel of Fig. \ref{fig:EnvelopeRotation} we present the velocity map in the plane $z=3 \times 10^{12} \cm$ at $t=2.25 P_{\rm orb}$, i.e., the same plane and time of the lower-left panel of Fig. \ref{fig:Dens_no_tilt}, but only of the inner region of this plane. Colour-coding gives the magnitude of the velocity and the arrows represent the velocity direction. We can note the global counterclockwise rotation of this region, although some small regions might have a clockwise sense of rotation.

We tentatively conclude that for the parameters that we use here, the influence of the jets on the envelope angular momentum component along the initial orbital angular momentum axis is smaller than that of the orbital angular momentum of the NS-RSG binary system, but it might be non-negligible. The effect strongly depends on the properties of the jets. In particular we recall that we expect the jets to be $\approx 10-50$ more energetic than what we use here (because of numerical limitations).  We encourage future CEE simulations with jets to launch jets in a scheme that conserves angular momentum and to record the angular momentum of the envelope 

We comment on the mass inside the numerical grid that we present in the lower panel of Fig. \ref{fig:J_components_and_mass}. 
Note that the mass is calculated inside the entire numerical grid, and not only in the shell inside which we calculate the envelope angular momentum. 
At early times the mass in the numerical grid increases because there is an inflow from the large area of the boundary of the numerical grid. Later the jet-induced outflowing gas exits the numerical grid and the mass decreases. 

\section{Angular momentum by tilted jets} 
\label{sec:Tilted}

\subsection{Tilting the envelope angular momentum} 
\label{subsec:Tilting}

We present the angular momentum that the tilted-jets deposit to the envelope by solid lines in the upper two panels of Fig. \ref{fig:J_components_and_mass}.
Again, by envelope we refer to the inflated envelope mass that is inside the shell $ 0.2 R_{\rm RSG} = 1.23 \times 10^{13} \cm < r < 2 R_{\rm RSG} = 1.25 \times 10^{14} \cm$. The component of the deposited angular momentum along the orbital angular momentum is similar, but not identical, in the two simulations we conduct here, $J_z({\rm tilt}) \approx J_z({\rm per})$ (middle panel of Fig. \ref{fig:J_components_and_mass}).  After the rapid rise in $J_z({\rm tilt})$ it fluctuates around its maximum value and then declines. Towards the end of our simulation $J_z({\rm tilt})$ levels off at a lower value. We expect that at later times it would rise again to have long-period fluctuations, similar to the fluctuations in $J_x({\rm tilt})$ and $J_y({\rm tilt})$ (upper panel). The fluctuation results from the complicated interaction of the jets with the highly-distorted envelope along the orbital motion.  

Unlike in the perpendicular-jets simulation, in the tilted-jets simulation the jets deposit angular momentum that has large $x$ and $y$ components. 
After one orbit we find the angular momentum that the jets deposited to the envelope to be $(J_x,J_y,J_z)_{\rm tilt,1} \simeq (2,4.5,20) \times 10^{52} \g \cm^2 \s^{-1}$, while after three orbits we find $(J_x,J_y,J_z)_{\rm tilt,3} \simeq (2.6,4,11) \times 10^{52} \g \cm^2 \s^{-1}$. This by itself implies that the average angular momentum of the envelope after three orbits is tilted by an angle of 
\begin{equation}
 \alpha_{\rm J,env}=\tan^{-1} 
 \left( \frac{\sqrt{J^2_x+J^2_y}}{J_z}  \right)_{\rm tilt}
 \simeq 23^\circ
 \label{eq:tileangles}   
\end{equation}
relative to the orbital angular momentum. We note, however, that the spiralling-in NS deposits even more angular momentum to the envelope than what the jets do for the parameters we use here, and so the tilted of the envelope angular momentum to the orbital angular momentum might be smaller than what we give in equation (\ref{eq:tileangles}). On the other hand, we expect the NS to launch stronger jets than what we use here because of numerical limitations.   

The initial moment of inertia of the envelope (from the inert core to the stellar radius) is $I_{\rm env,0}= 1.2 \times 10^{61} \g \cm^2$. The total deposited angular momentum after three orbits is $J({\rm tilt,3})\simeq 1.2\times 10^{53} \g \cm^2 \s^{-1}$. Had the envelope maintained its initial structure the spin period of the envelope would have been $P_{\rm spin,3} \approx 20 \yr$. The orbital period of a test particle on the surface of the undisturbed RSG star is $P_{\rm Kep,0}=2.3 \yr$. For the inflated envelope the radius is larger, but the envelope mass is lower. 
The jets by themselves can spin-up the giant envelope to $\approx 10\%$ of its breakup rotation speed. We do not perform a more accurate derivation because the envelope has lost any symmetry.
    
We go back to our numerical setting of the jets' power. We use here $\zeta=1.2 \times 10^{-4}$ in equation (\ref{eq:JetsPower}). We noted there that a more realistic value is $\zeta \simeq 0.002 - 0.005$ as we found in our earlier paper \citep{Hilleletal2022FB}. Namely, the jets might be as $\simeq 10-50$ as powerful as we can allow for numerical reasons in our simulations. 

We conclude that jets that a NS launches in a CEE might spin-up the RSG envelope to $> 0.1$ times the break-up velocity of the envelope. Although this rotation by itself might have a small direct effect on the envelope and the mass loss process, there might be indirect effects. One is the powering of a dynamo that substantially amplifies magnetic fields in the envelope. The second might come later, if the left-over of the envelope collapses onto the core of the RSG. In that case the envelope spins-up the remnant that later explodes as a stripped-envelope CCSN. The newly-born NS (or BH) remnant might have a final spin that is somewhat misaligned to the orbital angular momentum in case of tilted jets \citep{Soker2022misalignment}. The interaction of the NS with its companion tilts also the spin of the old NS that was spiralling-in inside the envelope. 

Overall, we strengthen the conclusion of \cite{Soker2022misalignment} that a CEE in a triple star system might lead to spin-orbit misalignment of NS/BH–NS/BH binary systems.  

\subsection{Comparison with analytical estimates. } 
\label{subsec:Comparison}

\cite{Soker2022misalignment} makes some simple assumptions to allow the calculation of the angular momentum that the jets deposit to the envelope. He assumes that the tilted jet that points out removes mass from the envelope along the jet's axis and that this gas carries all its angular momentum out, while the tilted jet that points into the star deposits all its angular momentum to the envelope. With the orientation we have here the assumption that jets remove mass along their axis implies that the jets deposit only $J_y$ component to the envelope. However, we saw above (e.g., Fig. \ref{fig:envelope_outflow_perp}) that the jets remove envelope mass not along their axis. Therefore, we can compare only the value of $J_y({\rm tilt})$ to the analytical estimates of \cite{Soker2022misalignment}. 

Under his simple assumptions, \cite{Soker2022misalignment} analytically estimates that the angular momentum the jets deposit to the envelope is 
\begin{equation}
d J_{y,{\rm a}} \approx \frac{2}{\pi} \cos \beta \sin \beta a v_{\rm esc} dM_{\rm env} ,
\label{eq:dJ2j}
\end{equation}
where $dM_{\rm env}$ is the mass that the jets remove from the envelope, $\beta$ is the angle of the jets relative to the equatorial plane (Fig. \ref{fig:Schamnatic}), $v_{\rm esc}$ is the escape velocity from the common envelope and $a$ is the orbital radius.
Substituting the values we use here, $\beta=45^\circ$, $v_{\rm esc} = 55 \km \s^{-1}$, and $a=700R_\odot=4.9 \times 10^{13} \cm$, in the tilted-jets simulation we find 
 \begin{equation}
\Gamma_{y,{\rm a}} \equiv \frac {d J_{y,{\rm a}} }{dM_{\rm env}}  \approx 8.5 \times 10^{19} \cm^2 \s^{-1} = 57 \AU \km \s^{-1} .  
\label{eq:Gamma}
\end{equation}
From our numerical results for the tilted-jets simulation we find the values after one, two, and three orbits to be $\Gamma_1 \simeq 30$, $\Gamma_2 \simeq 13$, and $\Gamma_3 \simeq 5.4 \AU \km \s^{-1}$.

Based on the ratio $\Gamma_{y,{\rm a}}/\Gamma_1 < 2$ we conclude that the angular momentum deposition in the first orbit and the analytical estimate are compatible with each other. However, at later times the deposition of angular momentum decreases. We further discuss these results in section \ref{sec:Reliability}.  

\section{On the reliability and implications of our simulations} 
\label{sec:Reliability}

There are a few drawbacks of our simulations. We list the main three as follows. (1) We do not follow the gravitational interaction of the NS with the envelope, nor the self gravity of the envelope. We do include the spherically symmetric gravitational field of the RSG as it is at $t=0$. (2) We assume a constant orbit and do not allow for in-spiralling as should be in a CEE. We do so in order to examine only the effects of the jets. In a forthcoming paper we will allow spiralling-in. (3) We start the simulations with the NS already inside the envelope. There is a partial justification for that besides limited numerical resources. In case the powering of the jets is by the disruption of a main sequence close companion to the NS (or BH), as we schematically present in the lower panel of Fig. \ref{fig:Schamnatic}, the strong jets will start to operate only when the NS/BH is inside the envelope. They will be active for a limited amount of time, possibly much less than one orbit, until the NS consumes the accretion disk.   

Despite these drawbacks, we argue that for the effects we study here our simulations do present the qualitatively correct results and quantitatively accurate to better than about an order of magnitude. This is because during the rapid spiralling-in (plunge-in) phase of the CEE the NS rapidly dives from the surface of the giant to deep inside the envelope. This implies that before the companion, here a NS, completes an orbit it dives deep into the envelope. Therefore to each new radius that the NS arrives it is the first orbit in that radius. Namely, the NS `starts' its evolution there. 

We argue therefore, that the increase in the value of the envelope angular momentum that we find in the first orbit (two upper panels of Fig. \ref{fig:J_components_and_mass}), from $t=0$ to $t=1.78 \yr$, is a real effect. We do not actually need to wait for the flow to  relax because during the plunge-in phase of the CEE the system reaches no relaxation. 
In addition, the $y$ component of the angular momentum that the jets deposit in the first orbit, $J_y({\rm tilt,1})$, is similar to the analytically estimated value (equations \ref{eq:dJ2j} and \ref{eq:Gamma}). 

After the plunge-in phase the system enters the self-regulated phase where the spiralling-in is very slow (e.g., \citealt{GlanzPerets2021}).
Our assumption of a constant orbital radius holds for this phase, although we take a large orbital radius of $a=700 R_\odot=4.9 \times 10^{13} \cm$ while the self-regulated phase will take place with $a < 100 R_\odot$.
 Therefore, our finding does not result as a numerical effect of the non-spiralling-in assumption but rather it is a real effect as our assumed orbital motion holds for the self-regulated phase. 
For that, we expect our finding that after about two orbits the rate of angular momentum deposition by the jets substantially decreases to hold during the self-regulated phase of the CEE. 
We therefore expect that during the self-regulated phase of the CEE not only the binary system deposition of angular momentum almost vanishes, but also the deposition of angular momentum by jets. 


\section{Summary} 
\label{sec:Summary}

We conducted exploratory numerical simulations to study the angular momentum that jets launched by a NS deposit to the common envelope. Because the interaction of the jets with the RSG envelope demands high computer resources, we omitted some important processes. We did not allow for spiralling-in, we did not include the self-gravity of the envelope (but included the RSG gravity), and we started with the NS already inside the envelope (section \ref{sec:Numerical}). 

Despite these drawbacks, we argued in section \ref{sec:Reliability} that our results do reproduce the qualitative results and to better than an order of magnitude accuracy the quantitative values. In particular, the angular momentum deposition during the first orbit of our numerical simulations represents the behavior during the plunge-in phase of the CEE, namely, during the phase when the compact companion rapidly spirals-in from the giant's surface to within deep inside its envelope. The second and third orbits represent the following self-regulated phase of the CEE when spiralling-in proceeds very slowly. We could not continue the simulations beyond the third orbit because they proceeded very slowly at those times.  

Our main goal is to explore the tilted angular momentum components that tilted jets deposit to the envelope, where by tilted we refer to jets with their symmetry axis inclined to the orbital angular momentum axis (Fig. \ref{fig:Schamnatic}).
This situation might occur in triple-star systems when a tight (inner) binary system with an inclined orbital plane enters the giant's envelope \citep{Soker2022misalignment}. We conduct one perpendicular-jets simulation where the jets are not tilted, i.e., the jets' axis is perpendicular to the equatorial plane ($\beta=90^\circ$), and one tilted-jets simulation where we set $\beta=45^\circ$.    

Because of numerical limitations we simulated jets that are only $\simeq 0.02-0.05$ times the power of the jets that we expect a NS to launch for the parameters of the simulations, i.e., orbit and envelope density at that orbit. Specifically, we took $\zeta=1.2 \times 10^{-4}$ in equation (\ref{eq:JetsPower}) instead of the expected values of $\zeta \simeq 0.002 - 0.005$ (\citealt{Hilleletal2022FB}; section \ref{subsec:Tilting}).

In section \ref{sec:FlowStructure} we discussed the complicated outflow structure that the jets induce, as we present in Figs. \ref{fig:Dens_no_tilt}, \ref{fig:Dens_45deg_tilt}, \ref{fig:rho_v_XY_YZ_tilt}, \ref{fig:envelope_outflow_perp}, and \ref{fig:EnvelopeRotation}.
Our main findings, however, refer to the angular momentum that the jets deposit to the envelope and are as follows.  

(1) \textit{Angular momentum component parallel to the orbital angular momentum.} 
Even jets that are not tilted (the perpendicular-jets simulation) deposit angular momentum to the envelope, but only along the orbital angular momentum (Two upper panels of Fig. \ref{fig:J_components_and_mass}). Although the value is only $J_z  ({\rm per}) \simeq 0.25 J_z ({\rm orb})$, where  $J_z ({\rm orb})$ is the binary orbital angular momentum at the onset of the CEE, we recall that in realistic scenarios we expect stronger jets than what we launch in our simulations. The tilted jets deposit a similar angular momentum along the orbital angular momentum axis, but with some time variations. The reason for this deposition of angular momentum is that during the first orbit the jets eject envelope mass with negative angular momentum (Fig. \ref{fig:envelope_outflow_perp}). We conclude that during the plunge-in phase jets that a NS/BH launches can substantially add to the angular momentum that the spiralling-in NS/BH deposits as it loses orbital angular momentum while spiralling-in. As far as we know, our numerical finding that jets can substantially spin-up the envelope along the orbital angular momentum is new.  

(2) \textit{Angular momentum component perpendicular to the orbital angular momentum.} Jets that are perpendicular to the orbital plane (aligned with the orbital angular momentum) do not deposit angular momentum with components perpendicular to the orbital angular momentum (dashed lines in the upper panel of Fig. \ref{fig:J_components_and_mass}). However, the angular momentum that the tilted jets deposit to the envelope has large $J_x({\rm tilted})$ and $J_y({\rm tilted})$ components (solid lines in the upper panel of Fig. \ref{fig:J_components_and_mass}). The jet-deposited angular momentum component $J_y({\rm tilted})$, i.e., that is perpendicular to both to the tilted-jets' axis and to the orbital angular momentum axis, that we find here is compatible with the analytical estimate by  \cite{Soker2022misalignment}. The values that we find here are similar to the analytical estimate (equations \ref{eq:dJ2j} and \ref{eq:Gamma}). However, that the jets deposit angular momentum with a large  $J_x({\rm tilted})$ component is new. 

We encourage future simulations of CEE with jets and of the grazing envelope evolution to follow the angular momentum that the jets deposit to the envelope. 

The general evolution we consider here and the tilting of the envelope angular velocity by tilted jets have two implications. The first one is on the outflow morphology. Triple-star systems where the inner (of the close binary) orbital plane is inclined to the outer (triple) orbital plane lead to the ejection of `messy' nebulae, like messy planetary nebulae (e.g., \citealt{BearSoker2017}). Namely, nebulae that have no symmetry at all, neither mirror symmetry, nor axisymmetry, nor point-symmetry. An inclined envelope rotation might add to the departure of the outflow from any kind of symmetry by its influence on the wind from the common envelope, for example by inducing magnetic activity with inclined dipole. 

The second implication of tilted jets is to the remnants. The interaction of the old NS/BH (the one that enters the RSG envelope) with its close companion tilts its spin. This implies that this NS/BH will have a spin misaligned with the orbital angular momentum. Our results imply that the newly born NS/BH, the one that is formed after the CEE in a CCSN explosion of the core, might also have a misaligned spin, as \citep{Soker2022misalignment} suggested from analytical estimates. This will be the case if the final mass of the envelope collapses onto the core of the RSG.

We therefore confirm and put on a more solid ground the claim  \citep{Soker2022misalignment} that a CEE in a triple star system might lead to spin-orbit misalignment of NS/BH–NS/BH binary systems.

\section*{Acknowledgments}

 We thank an anonymous referee for helpful comments and suggestions.  
This research was supported by the Amnon Pazy Research Foundation.

\section*{Data availability}
The data underlying this article will be shared on reasonable request to the corresponding author.  


\label{lastpage}

\end{document}